%% file: main.tex
\documentclass[sigconf, nonacm]{acmart}

\AtBeginDocument{%
  }

\usepackage{multirow}
\usepackage{tabularx}
\usepackage{placeins}
\usepackage{enumitem}  
\usepackage{algorithmic}
\usepackage{algorithm}
\usepackage{dblfloatfix}
\usepackage{float}
\usepackage[most]{tcolorbox}
\usepackage{makecell}

\begin{document}

\title[U-Lens]{U-Lens: Supporting User Uncertainty Management in Long-Form LLM Responses}

%
\author{Yu Mei}
\affiliation{
  \institution{Tsinghua University}
  \city{Beijing}
  \country{China}
}
\email{meiy24@mails.tsinghua.edu.cn}

\author{Qingyue Zhuang}
\affiliation{
  \institution{Tsinghua University}
  \city{Beijing}
  \country{China}
}
\email{zhuangqy23@mails.tsinghua.edu.cn}

\author{Jie Cai}
\affiliation{
  \institution{Tsinghua University}
  \city{Beijing}
  \country{China}
}
\email{jie-cai@mail.tsinghua.edu.cn}

\author{Chang Liu}
\affiliation{
  \institution{Tsinghua University}
  \city{Beijing}
  \country{China}
}
\email{c-liu21@tsinghua.org.cn}

\author{Zhi Zheng}
\affiliation{
  \institution{Tsinghua University}
  \city{Beijing}
  \country{China}
}
\email{zhengz22@mails.tsinghua.edu.cn}

\author{Zhoutong Ye}
\affiliation{
  \institution{Tsinghua University}
  \city{Beijing}
  \country{China}
}
\email{yezt24@mails.tsinghua.edu.cn}

\author{Chun Yu}
\affiliation{
  \institution{Tsinghua University}
  \city{Beijing}
  \country{China}
}
\email{chunyu@tsinghua.edu.cn}

\author{Yuanchun Shi}
\affiliation{
  \institution{Tsinghua University}
  \city{Beijing}
  \country{China}
}
\email{shiyc@tsinghua.edu.cn}

\renewcommand{\shortauthors}{Mei et al.}

\begin{abstract}

\input{data/abstract}
\end{abstract}

\ccsdesc[500]{Human-centered computing~Interactive systems and tools}

\keywords{Uncertainty management, Large language models, Human-AI interaction}


\maketitle

\input{data/chap1}
\input{data/chap2}

\input{data/chap3}
\input{data/chap4}
\input{data/chap5}

\input{data/chap6}
\input{data/chap7}
\input{data/chap8}

\bibliographystyle{ACM-Reference-Format}
\bibliography{ref/main_citation}

\clearpage
\appendix
\input{data/appendix}
\end{document}

%% file: data/abstract.tex
Uncertainty can appear throughout LLM-generated text (e.g., questionable claims, ambiguous terms). Prior work largely focuses on making such uncertainty visible through cues such as confidence scores, but seeing uncertainty is not the same as managing it. Through a formative study, we examine uncertainty management across interpretation, evaluation, and decision. From these insights, we derive design guidelines for uncertainty target representation, evaluative explanation, response guidance, and interactive presentation. We instantiate them in U-Lens, an uncertainty-management system that organizes uncertain information into contextual inspection targets, prioritizes them, and links each to evaluative context and response options. In an 18-participant within-subjects study comparing U-Lens with a confidence-cue baseline, U-Lens improved verification efficiency and effort allocation, reduced perceived workload, and strengthened support across all three stages. This work reframes uncertainty support for generative AI from text-centered cues to a user-centered process of interpreting, evaluating, and acting on uncertainty.

%% file: data/chap1.tex
\section{Introduction}
Large language models (LLMs) are increasingly used to generate long-form answers for knowledge-intensive tasks, such as learning unfamiliar topics, synthesizing background information, drafting analytical reports, and supporting decision-making \cite{dhillon2024shaping, yun2025generative, spatharioti2025effects}. In these contexts, users must work through extended responses that may contain factually questionable claims, vague or underspecified explanations, unfamiliar terminology, and statements whose relevance or reliability is difficult to judge \cite{subramonyam2024bridging, gero2024supporting}. Because these sources of uncertainty are often embedded within otherwise plausible and coherent text, users may struggle to decide which parts of a response deserve scrutiny, why they may be unreliable, and what to do next \cite{kalai2025language, cheng2024relic}.

Prior work has explored ways of presenting uncertainty cues to users in AI and LLM outputs, including confidence scores, uncertainty markers, visual cues, and natural-language explanations \cite{xu2025confronting, prabhudesai2023understanding, schmidt2024natural}. These approaches are valuable for making uncertainty visible. However, seeing uncertainty is not the same as managing it. A cue can draw attention to possible unreliability, but users still need to locate what the concern is about, interpret why it matters, and decide whether and how to act on it. This challenge becomes especially pronounced in long-form responses, where uncertainty can be distributed across many claims and relations within a complex text.

We refer to this broader process as \textit{user uncertainty management}: the user-centered process of interpreting, evaluating, and acting on uncertainty-relevant content. In this paper, we investigate how interactive systems can support user uncertainty management in long-form LLM responses.

To investigate this problem, we pose four research questions:
\begin{itemize}
    \item \textbf{RQ1}: How do users manage uncertainty in long-form LLM responses?
    \item \textbf{RQ2}: What design guidelines can support user uncertainty management in long-form LLM responses?
    \item \textbf{RQ3}: How can these design guidelines be instantiated in an interactive uncertainty management support system?
    \item \textbf{RQ4}: How does such a system support uncertainty management in long-form LLM response verification?
\end{itemize}

We first conducted a formative study to examine how users manage uncertainty in long-form LLM responses and what support they need during this process. Drawing on the Theory of Motivated Information Management (TMIM) \cite{afifi2004toward}, we organized our analysis around three stages of uncertainty management: interpretation, evaluation, and decision. Based on these insights, we derive design guidelines that address both stage-specific and cross-stage needs (DG1--DG4): uncertainty target representation, evaluative explanation, response guidance, and interactive presentation.

Building on these design guidelines, we introduce U-Lens, an uncertainty-management support system for long-form LLM responses. Rather than presenting uncertainty as a collection of detached model signals, U-Lens organizes uncertainty-relevant content into contextual inspection targets, prioritizes these targets for attention, and connects each target with evaluative context and response options. 

We evaluated U-Lens through a controlled within-subjects study with 18 participants, comparing it against a confidence-cue baseline in a limited-budget handoff verification task. Our results show that U-Lens increased verification efficiency, improved users' allocation of verification effort, and lowered perceived task workload. It also provided stronger perceived support across interpretation, evaluation, and decision stages.

This study contributes to research on human-AI information processing by extending and operationalizing TMIM for AI-generated information. Building on this process view, we move uncertainty support beyond presenting isolated, text-centered signals toward scaffolding how users interpret, evaluate, and act on uncertain generated content. More broadly, we argue for shifting generative AI interface design from supporting the answer itself to supporting the user's work with the answer, making user sensemaking a critical design goal alongside content generation and evidence provision.


%% file: data/chap2.tex
\section{Related Work}

\subsection{Uncertainty in Long-Form LLM Responses}
Long-form LLM responses are increasingly used for learning, writing, and decision-making, yet they often mix supported statements with unsupported details, vague explanations, and inconsistent relations \cite{massenon2025my, alansari2026large}. Prior factuality benchmarks show that errors in long-form generation are distributed across many atomic facts rather than captured by a single response-level judgment \cite{min2023factscore, wei2024long}. This makes uncertainty difficult for users to handle in long-form LLM responses.

A large body of work aims to reduce factual errors and unsupported content in LLM outputs. Some methods improve factual grounding during generation or alignment \cite{tian2023fine, zhang2024self, lin2024flame}; others encourage models to express uncertainty in long-form answers \cite{yang2025logu, band2024linguistic, zhang2026lovec}, check their own outputs during generation \cite{dhuliawala2024chain, asai2024self}, or retrieve external evidence to revise unsupported content after generation \cite{gao2023rarr, gou2023critic, li2024rac}. These approaches provide important resources for making generated text more reliable, but their primary goal is usually to improve model output rather than helping users make sense of the output uncertainty itself.

Another line of work estimates uncertainty in LLM outputs \cite{xia2025survey, shorinwa2025survey}. Existing methods include verbalized uncertainty, where models express confidence through natural language or numeric scores \cite{lin2022teaching, tian2023just, xiong2023can}; token-level uncertainty, which uses generation signals such as token probabilities, log-likelihood, and entropy \cite{jiang2021can, manakul2023selfcheckgpt, zhang2023enhancing, wightman2023strength}; consistency-based uncertainty, which measures agreement across sampled outputs and can be extended with semantic clustering \cite{cole2023selectively, lyu2025calibrating, hou2023decomposing, kuhn2023semantic, farquhar2024detecting, nikitin2024kernel}; and mechanistic interpretability, which examines internal mechanisms associated with model uncertainty \cite{belinkov2022probing, cunningham2023sparse, gurnee2023finding}. These approaches provide useful signals for assessing LLM reliability, but they primarily treat uncertainty as a model-side signal to be estimated or analyzed. They do not fully address how everyday users should interpret many uncertainty signals distributed across a long response.

More recent interactive systems make reliability evidence more inspectable for users. RELIC, for example, uses self-consistency across multiple long-form samples to help users assess generated claims \cite{cheng2024relic}. Other systems present source attributions, factuality scores, or visual factuality indicators so that users can inspect the evidence behind generated content \cite{gao2023enabling, do2024facilitating, do2025highlight}. These systems show the value of exposing reliability-related evidence during LLM interaction. However, they still leave open a broader interaction problem: how users should manage many possible uncertainty targets and signals while reading a long response, especially when they cannot inspect or verify everything.

Our work shifts the focus to supporting the broader process of user uncertainty management. U-Lens uses model-side signals and target-level evidence as inputs, but organizes them into manageable interaction objects for user-facing support.

\subsection{Human-Centered Uncertainty Support in LLM Outputs}
Communicating uncertainty in model outputs is important across many everyday use scenarios, including learning \cite{sanchez2022deep}, writing \cite{li2024value, poddar2023ai}, healthcare \cite{toh2025effect}, decision-making \cite{cao2024designing}, and information seeking \cite{kim2024m}. Prior work has externalized uncertainty through different forms, including numerical forms such as probabilities and confidence scores \cite{lin2022teaching, zhang2020effect, yang2024verbalized}, graphical forms such as visual encodings of distribution or ambiguity \cite{do2025highlight, spatharioti2025effects}, and verbalized forms that communicate uncertainty through natural language \cite{bhatt2021uncertainty, xiong2023can, zhou2024relying}. These forms can help users notice potentially unreliable information, calibrate their reliance, and decide whether further checking is needed \cite{zhou2023synthetic, jakesch2023co, schoeffer2024explanations}.

In LLM-based interaction, verbalized forms of uncertainty communication are especially salient because users already seek information, interpret answers, and make decisions through language. Such expressions may also align with familiar patterns of human communication, making uncertainty cues intuitive for users to interpret \cite{ulmer2025anthropomimetic, xiong2023can}. 

Verbalized uncertainty can appear as hedges, confidence statements, uncertainty explanations, or textual cues about what may be unreliable \cite{lin2022teaching, xiong2023can, zhou2024relying}. Prior work shows that such expressions can shape users' trust \cite{xu2025confronting}, confidence \cite{kim2024m}, perceived reliability or accuracy \cite{spatharioti2025effects, steyvers2025large}, reliance \cite{kim2025fostering}, agreement with model outputs \cite{kim2024m}, decision-making \cite{xu2025confronting}, and task performance \cite{kim2024m, xu2025confronting}. This makes verbalized uncertainty an important design resource for human-centered uncertainty support.

Existing studies have examined several aspects of verbalized uncertainty design. Some work focuses on linguistic choices, such as how strongly uncertainty is expressed \cite{zhou2023navigating, zhou2024relying, yona2024can, liu2025metafaith}, whose perspective the uncertainty is voiced from \cite{kim2024m, zhou2024relying}, and what textual scope the cue is attached to \cite{do2024facilitating, yona2024can}. Other work examines presentation choices, such as when uncertainty cues are shown and how prominently they are displayed \cite{kim2024m, do2025hide}. Researchers have also explored contextual grounding around uncertainty cues, including factuality indicators \cite{do2025highlight}, source attributions \cite{do2024facilitating, ding2025citations}, and inconsistency cues \cite{kim2025fostering}.

However, much of this work treats uncertainty support as the design of individual cues at specific textual points or interaction moments. Such cues are useful for noticing uncertainty, but they can leave users with a broader management problem: in a long response, users still need to decide which uncertain parts deserve attention, how these parts relate to the task, and what to do next under limited time and effort. In this work, we study how systems can support users' continuous process of localizing, evaluating, and responding to uncertainty in long-form LLM responses.

\subsection{From Explainable AI to User Uncertainty Management}
HCI research has shown that transparency and explanations can help users understand, question, and appropriately rely on AI systems \cite{lim2009and, buccinca2021trust, vasconcelos2023explanations}. Much of this work focuses on explaining how an AI system reaches a specific decision, recommendation, or classification: why the system produced a particular output \cite{lim2009and, liao2020questioning, lim2009assessing}, how that output might change under different conditions \cite{lim2010toolkit, liao2020questioning}, and which inputs, features, contextual cues, or examples shaped the result \cite{hohman2019gamut, kaur2020interpreting, hong2020human}. These explanations are especially useful when users need to understand an AI-mediated decision process, such as why a system produced a recommendation, classification, or risk judgment.

In this paper, we focus on a different object: uncertainty embedded in the content produced by generative models. Unlike uncertainty in a bounded system decision, uncertainty in long-form generative outputs is distributed across the generated text itself. A response may contain uncertain claims, vague statements, unsupported details, and questionable relations at different textual granularities. The user's problem is therefore not only to understand why the system produced an output, but to decide which parts of the output deserve attention, how those parts should be evaluated, and what to do next.

This distinction shifts uncertainty support from explaining a system decision to supporting a broader process of uncertainty management. Prior work has shown that explanations can support calibrated trust and appropriate reliance \cite{yin2019understanding, zhang2020effect, buccinca2021trust}, shaping whether users accept, reject, verify, or override AI recommendations \cite{chen2023understanding, vasconcelos2023explanations, schoeffer2024explanations}. In long-form LLM responses, however, users face multiple localized and interdependent uncertainties. They need support not only for understanding a single output, but also for the ongoing process of localizing, evaluating, and responding to uncertainty.

To frame this broader process of user uncertainty management, we draw on the Theory of Motivated Information Management (TMIM). TMIM explains uncertainty management as a motivated process across interpretation, evaluation, and decision stages \cite{afifi2004toward}. Prior TMIM research has examined uncertainty management in interpersonal, relational, and health contexts \cite{afifi2006seeking, afifi2006examining, afifi2009avoidance}, and later extended this account to technology-mediated settings such as online information seeking, online support groups, and social media \cite{li2020applying, kanter2019use, seo2022international}. Prior work has also used TMIM's three-stage structure to explain how people seek information \cite{afifi2006seeking}, manage misinformation \cite{sathianathan2025general}, navigate relationship uncertainty \cite{tokunaga2014seeking}, and engage with online social media \cite{kuang2022offline}.

These applications make TMIM a useful lens for our setting because they frame uncertainty management as interpreting and acting on textual information, such as social media posts and online health discussions. Long-form LLM responses pose a related but distinct challenge: users must manage uncertainty in AI-generated content whose reliability can vary across details, claims, and relations. We therefore build on TMIM to examine how interaction support can help users localize, evaluate, and respond to uncertainty in long-form LLM responses.

%% file: data/chap3.tex
\section{Formative Study}
We conducted a formative study to understand how users make sense of and respond to uncertainty they perceive in long-form LLM responses, and what additional information they need to support this process. Drawing on TMIM \cite{afifi2004toward}, we structure our findings around three stages of user uncertainty management---interpretation, evaluation, and decision---and derive design guidelines for supporting this process.

\subsection{Setup}
\subsubsection{Participants}
We recruited 12 participants (4 males, 8 females; ages $M=23.00, SD=2.04$) from social media. All participants reported daily use of LLMs across a range of activities, including information seeking (11/12), writing (11/12), programming (11/12), coursework and learning (10/12), data analysis (8/12), and entertainment (6/12). Each participant received \$10 upon completing the study. Demographic details are provided in Appendix~\ref{section: Participant Demographics}.

\subsubsection{Study Materials}
To cover diverse knowledge-intensive long-form generation scenarios, we selected three representative task types commonly studied in prior work: biography generation \cite{min2023factscore, kang2023ever, lin2024flame, yang2025logu, zhao2024wildhallucinations}, event explanation \cite{lin2024flame, wei2024long, yang2025logu, zhao2024wildhallucinations}, and scientific explanation \cite{lin2024flame, wei2024long, yang2025logu, zhao2024wildhallucinations}.

We first constructed a candidate pool from existing resources, using the English Wikipedia People Dataset \footnote{\url{https://www.kaggle.com/datasets/wikimedia-foundation/english-wikipedia-people-dataset}} for person entities and the LongFact dataset \cite{wei2024long} for historical events and scientific concepts. We collected 30 candidate entities for each task type. 
We then conducted a pilot screening with 10 participants, who rated each entity on topic familiarity and knowledge of its details (5-point Likert scale). We retained entities that were recognizable to a majority of pilot participants ($7/10$ or more) but not well known in detail ($M_{knowledge}<2.5$), from which we selected 8 final entities per task type. This ensured participants could make plausibility judgments without relying primarily on prior knowledge.

For each final entity, we used Qwen3-8B \footnote{\url{https://huggingface.co/Qwen/Qwen3-8B}}, one of the most widely used open-source LLMs, to generate the long-form outputs shown to participants. Its generation quality was sufficient for supporting our study goals. Examples of the generated materials are provided in Appendix~\ref{section: Study Materials}.


\begin{figure*}[t] 
\centering
\includegraphics[width=0.8\textwidth]{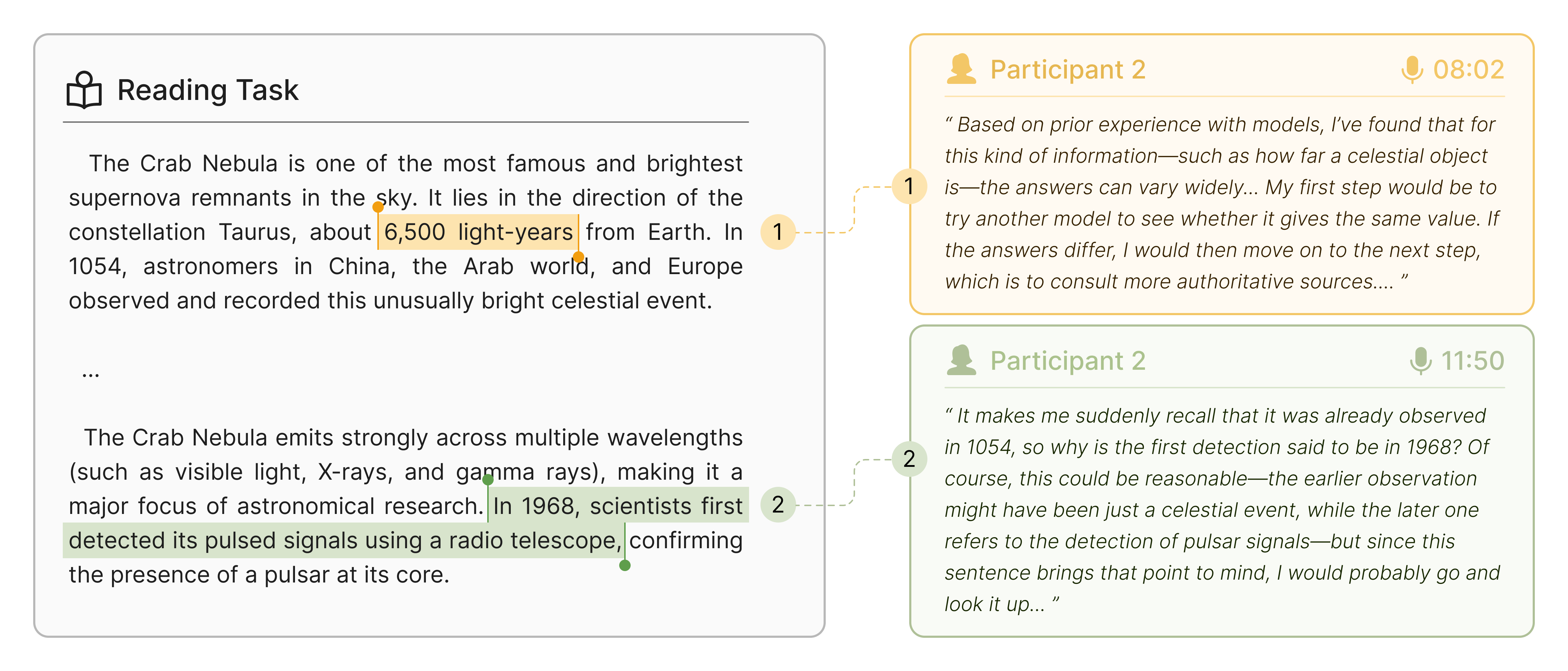}
\caption{Example of the think-aloud reading task. Participants highlighted text spans they perceived as uncertain from long-form LLM outputs, while explaining what made the span seem uncertain, how they made sense of that uncertainty, what they might do next, and what additional information would help them better understand or respond to it.}
\label{figure:formative_study_example}
\end{figure*}

\subsubsection{Procedure}
The study consisted of a think-aloud reading task followed by a semi-structured interview. 

In the reading task, each participant was randomly assigned one reading material from each of the three task categories. The presentation order of the three task categories was counterbalanced. While reading, participants were asked to freely highlight any claim or text span they perceived as uncertain, potentially inaccurate, or requiring further verification.
For each highlighted span, participants were asked to explain what made the span seem uncertain, how they made sense of that uncertainty, what they might do next, and what additional information would help them better understand or respond to it (see \autoref{figure:formative_study_example}). This process was repeated until they finished reading each material.

After completing the three reading tasks, participants took part in a semi-structured interview reflecting on their experiences. We asked about what made LLM-generated content seem uncertain, how they made sense of these uncertainties, what actions they would consider in response, and what kinds of system-provided information would help. We also probed which information needs were consistent across task types and which were task-dependent. 

All sessions were audio-recorded and transcribed. Two HCI researchers conducted thematic analysis \cite{clarke2017thematic}. We first open-coded all transcripts, then drew on TMIM \cite{afifi2004toward} to cluster and organize the codes into the final findings.

\subsection{Findings}
We found that participants' uncertainty-related practices align with the three-stage process described by TMIM \cite{afifi2004toward}: interpretation, evaluation, and decision. Accordingly, we organize our findings around how participants localized potentially uncertain content in the text (Section~\ref{section: formative study findings 1}), evaluated what that uncertainty meant (Section~\ref{section: formative study findings 2}), and decided whether and how to respond (Section~\ref{section: formative study findings 3}). Beyond these stage-specific practices, we also identify a cross-stage need for uncertainty management support to be prioritized, layered, and adjustable (Section~\ref{section: formative study findings 4}).

\subsubsection{Users Localize Uncertainty Across Multiple Textual Structures}
\label{section: formative study findings 1}
Participants did not treat uncertainty as belonging to a single fixed textual unit (see \autoref{tab: textual structures of uncertainty}). Instead, they localized uncertainty across multiple textual structures, including unit-based structures such as tokens, sentences, and broader segments, as well as relation-based structures within or between sentences.

Although the salience of these textual structures varied across task types and participants, this pattern appeared across all three task types: biography generation, event explanation, and scientific explanation.

In addition, determining the appropriate location and granularity of uncertainty was itself ``uncertain''. During the study, participants reported difficulty in directly judging whether a piece of information should be considered uncertain. For instance, when reading the historical material, P2 noted, ``It was somewhat difficult to judge, because I was not sure whether I was actually uncertain about this part.'' Participants also hesitated over how narrowly or broadly to highlight uncertain spans, reflecting ambiguity in defining the granularity of uncertainty.

\begin{table*}[]
\small
\caption{Participants localized uncertainty across multiple textual structures, including unit-based and relation-based structures, rather than at a single fixed unit.}
\label{tab: textual structures of uncertainty}
\begin{tabular}{llll}
\toprule
                                                                           & \multicolumn{1}{c}{\textbf{Textual Structure}}                    & \multicolumn{1}{c}{\textbf{Definition}}                                                                                                                              & \multicolumn{1}{c}{\textbf{Example}}                                                                                                                                                  \\ \hline
\multirow{3}{*}{\begin{tabular}[c]{@{}l@{}}Unit-\\ based\end{tabular}}     & Token level                                                       & \begin{tabular}[c]{@{}l@{}}Uncertainty about a single word, phrase, \\ factual detail, or domain-specific term.\end{tabular}                                          & \begin{tabular}[c]{@{}l@{}}P7: ``The timing of it might be worth \\ checking to see whether it is incorrect.''\end{tabular}                                                           \\ \cline{2-4} 
                                                                           & Sentence level                                                    & \begin{tabular}[c]{@{}l@{}}Uncertainty about a complete statement\\ or proposition, where the user evaluates\\ the validity of the sentence as a whole.\end{tabular} & \begin{tabular}[c]{@{}l@{}}P1: ```This event marked...' I have doubts\\ about this statement.''\end{tabular}                                                                          \\ \cline{2-4} 
                                                                           & Segment level                                                     & \begin{tabular}[c]{@{}l@{}}Uncertainty about a broader passage\\ typically due to insufficient background\\ knowledge or lack of evaluative cues.\end{tabular}       & \begin{tabular}[c]{@{}l@{}}P3: ``For the observation history part, I'm not\\ clear about the background information, so \\ I would want to verify that whole paragraph.''\end{tabular} \\ \hline
\multirow{2}{*}{\begin{tabular}[c]{@{}l@{}}Relation-\\ based\end{tabular}} & \begin{tabular}[c]{@{}l@{}}Intra-sentence\\ relation\end{tabular} & \begin{tabular}[c]{@{}l@{}}Uncertainty about the relationships\\ among multiple pieces of information\\ within a single sentence.\end{tabular}                       & \begin{tabular}[c]{@{}l@{}}P2: ``He was of Italian ancestry and grew \\ up in San Francisco; I wonder whether the\\ combination of information is accurate.''\end{tabular}            \\ \cline{2-4} 
                                                                           & \begin{tabular}[c]{@{}l@{}}Inter-sentence\\ relation\end{tabular} & \begin{tabular}[c]{@{}l@{}}Uncertainty arising from logical \\ inconsistency between sentences.\end{tabular}                                                         & \begin{tabular}[c]{@{}l@{}}P2: ``It said earlier that it had already been\\ observed in 1054, so why does it then say it\\ was first detected in 1968?''\end{tabular}                 \\ \bottomrule
\end{tabular}
\end{table*}

\begin{table*}[htbp]
\small
\caption{Textual and user-side cues participants used to evaluate uncertainty.}
\label{tab:uncertainty_cues}
\centering
\begin{tabular}{p{2cm} p{3.5cm} p{7.5cm}}
\toprule
\textbf{Cue Source} & \textbf{Cue Type} & \textbf{Description} \\
\midrule
\multirow{4}{*}{Textual cue}
& Specific detail
& Exact dates, numbers, names, and titles often trigger suspicion. \\
\cmidrule(lr){2-3}
& Vagueness / underspecification
& Unclear reference, underspecified explanation, or vague wording makes content hard to trust. \\
\cmidrule(lr){2-3}
& Information overload
& Dense stacking of details increases the sense that some part of the text is wrong. \\
\cmidrule(lr){2-3}
& Logical inconsistency
& Timeline conflicts, causal mismatch, or incoherent use-case descriptions trigger doubt. \\
\midrule
\multirow{4}{*}{User-side prior}
& Lack of background knowledge
& Users become broadly uncertain when they lack enough knowledge to judge plausibility. \\
\cmidrule(lr){2-3}
& Conflict with prior knowledge
& Users become uncertain when the text conflicts with what they already know. \\
\cmidrule(lr){2-3}
& Model distrust
& Users rely on prior beliefs about where LLMs are likely to make mistakes. \\
\cmidrule(lr){2-3}
& Personal habit
& Individual reading habits and disciplinary sensitivities shape uncertainty judgments. \\
\bottomrule
\end{tabular}
\end{table*}

\subsubsection{Users Evaluate Uncertainty Through Textual Cues and Their Own Knowledge}
\label{section: formative study findings 2}
After localizing potential uncertainty, participants evaluated what that uncertainty meant by combining two sources of cues: signals from the text itself and their own prior knowledge (see \autoref{tab:uncertainty_cues}).

Participants evaluated uncertainty through textual cues. Highly specific details often made participants suspicious. Conversely, vague or under-explained content also raised doubts. Beyond individual words or facts, they also evaluated larger textual patterns, perceiving densely stacked details or inconsistencies as indicators of potential errors.

At the same time, participants' own knowledge shaped how they evaluated uncertainty. In domains where they lacked background knowledge, participants often experienced global uncertainty rather than questioning specific points. In contrast, when possessing partial knowledge, they identified conflicts between the text and what they already knew. Participants also brought prior beliefs about LLMs into evaluation, such as expectations that models may fabricate precise details or produce weakly grounded summaries. 

In the interviews, participants' expressed support needs reflected this dual evaluation process. For evaluating potential issues in the text, they wanted inspectable grounding, such as source links or mappings from generated claims to original text (P4: ``provide reference sources for this part''). They also wanted explanations of why a claim, relation, or passage might warrant scrutiny. As P5 noted, ``the LLM could explain why it is uncertain.'' Some participants also found it useful to compare multiple possible answers generated by the LLM (P12). 

For evaluating uncertainty relative to their own knowledge, participants wanted support that adapted to their familiarity with the domain. When they were unfamiliar with the topic, they needed background explanations before they could decide whether a claim was problematic or worth verifying. They suggested that the system could provide a side note introducing the background context (P11), or offer explanations through a pop-up window (P5). When dealing with highly specialized domains, participants also wanted the system to ``explain unfamiliar terms'' (P8) or ``provide more detailed explanations'' (P1).

\subsubsection{Users Decide Response Strategies Based on Uncertainty Structure, Cue, and Resolution Cost}
\label{section: formative study findings 3}
After evaluating uncertainty, participants decided whether and how to respond. They did not treat every uncertainty as an error that required immediate verification (P6: ``If it doesn’t significantly affect my understanding of the text, I can keep it in mind as uncertain without investigating it further.''). Instead, they chose among different response strategies, including web search, asking follow-up questions of the current model, cross-model comparison, deferring judgment, or abandoning uncertain content. These choices were shaped by where uncertainty was located in the text, what cues made it salient, and how costly participants expected it to be to resolve.

For highly specific factual information (e.g., names, dates, numbers, locations), participants most often turned to external search. These details not only triggered suspicion, but were also perceived as relatively easy to verify. For example, P1 explained that ``searching would be faster'' for checking a concrete factual detail.

In contrast, when uncertainty came from vagueness, under-explanation, unfamiliar terminology, or missing background, participants were more likely to continue querying the model. They wanted the model to clarify what a term meant, expand a vague statement, explain the logic behind a conclusion, or provide background context. In these cases, uncertainty was harder to resolve through browsing alone, whereas the LLM ``had the grounding'' (P7) to answer more directly.

When uncertainty extended beyond a local target to a broader segment, participants rarely attempted to verify every claim one by one. Instead, they shifted to coarser response strategies, such as asking the current model to regenerate or summarize the passage, using another model for comparison, or abandoning the passage altogether. 

Finally, participants selectively ignored or deferred some uncertainties, especially when the issue was low-stakes or costly to verify. As P6 noted, if a doubtful point did not significantly affect later understanding and was difficult to verify immediately, he could ``keep it in mind as uncertain'' without further investigative effort.

\subsubsection{Users Need Prioritized, Layered, and Adjustable Support Across the Uncertainty Management Process}
\label{section: formative study findings 4}
Across interpretation, evaluation, and decision stages, participants did not want uncertainty-related information to be presented all at once. Instead, they preferred prioritized, layered, and adjustable support, so that they could manage their attention and verification effort throughout the uncertainty management process.

A recurring need was prioritization. Participants wanted the system to surface the most uncertain or most worth-checking parts first, helping users allocate limited verification effort. Some suggested showing only low-confidence parts (P12: ``Parts with confidence maybe below 30 or 40 could be shown first''). Others wanted the system to first filter suspicious content and then let users inspect the details as needed (P9: ``The system can filter it for me first, and then I can check it as needed'').

Participants also preferred layered and on-demand presentation. For example, P7 suggested that the system could initially show a small amount of information to support rapid assessment, while allowing users to reveal additional details when necessary. Similarly, P5 suggested placing a small link or marker beside a sentence so that users could click to view details only when needed. P12 preferred confidence information to remain hidden until explicitly requested, because they did not need to inspect every sentence.

Finally, participants wanted adjustable control over the scope and amount of support. Some proposed setting personalized confidence thresholds, so that the system would automatically highlight content whenever ``its confidence falls below a certain level'' (P10). Others suggested adjusting the level of displayed support through user prompts during interaction, for example by telling the system when its current response was ``too detailed or too brief'' (P11).

\subsection{Design Guidelines for Supporting User Uncertainty Management in Long-Form LLM Responses}
\label{section: design guidelines}
Drawing on the formative findings, we derive design guidelines for supporting user uncertainty management in long-form LLM responses (\autoref{tab:design_guidelines}). These guidelines translate the empirical findings into four forms of support that correspond to the three stages and cross-stage needs identified above---uncertainty target representation (DG1), evaluative explanation (DG2), response guidance (DG3), and interactive presentation (DG4).

\begin{table*}[htbp]
\small
\centering
\caption{Design guidelines for supporting user uncertainty management in long-form LLM outputs.}
\label{tab:design_guidelines}
\begin{tabular}{p{2.0cm} p{1.5cm} p{3.2cm} p{5.5cm}}
\toprule
\textbf{} 
& \textbf{TMIM Stage} 
& \textbf{Design Guideline} 
& \textbf{Support Strategies} \\
\midrule

DG1. Target representation 
& Interpretation 
& Help users localize uncertainty across multiple textual structures.
& Unit-based (token level, sentence level, segment level); relation-based (intra-sentence relation, inter-sentence relation)  \\

\midrule

DG2. Evaluative explanation 
& Evaluation 
& Help users evaluate uncertainty by explaining both text-oriented cues and user-relevant knowledge needs. 
& Text-oriented explanation (wording, source grounding, uncertainty rationale, alternative answers); user-aligned explanation (background context, term explanation, adjustable detail) \\

\midrule

DG3. Response guidance
& Decision 
& Help users choose whether and how to respond based on uncertainty structure, cue, and resolution cost.
& Response actions (web search, same-model query, cross-model comparison, content discard, judgment deferral); selection criteria (textual structure, uncertainty cue, resolution cost) \\

\midrule

DG4. Interactive presentation
& Cross-stage 
& Help users manage attention and verification effort through prioritized, layered, and adjustable presentation of uncertainty information.
& Priority-based filtering and ranking; lightweight initial cues; progressive and on-demand disclosure; adjustable control (personalized confidence thresholds, adjustable level of detail) \\

\bottomrule
\end{tabular}
\end{table*}

%% file: data/chap4.tex
\section{Design and Implementation}
\label{section: DESIGN AND IMPLEMENTATION}
We present U-Lens, an interactive uncertainty management support system for long-form LLM responses. U-Lens instantiates the design guidelines from Section~\ref{section: design guidelines} through three coordinated interface regions: an \textit{LLM Interaction View}, where users submit queries to an LLM and inspect uncertainty targets within the context of the generated response (DG1); a \textit{Prioritized Uncertainty Panel}, which ranks and filters uncertainty targets through adjustable, user-aligned controls (DG4); and an \textit{Uncertainty Detail Card}, which presents explanations and response guidance for a selected uncertainty target (DG2-DG3).

\begin{figure*}[htb] 
\centering
\includegraphics[width=1.0\textwidth]{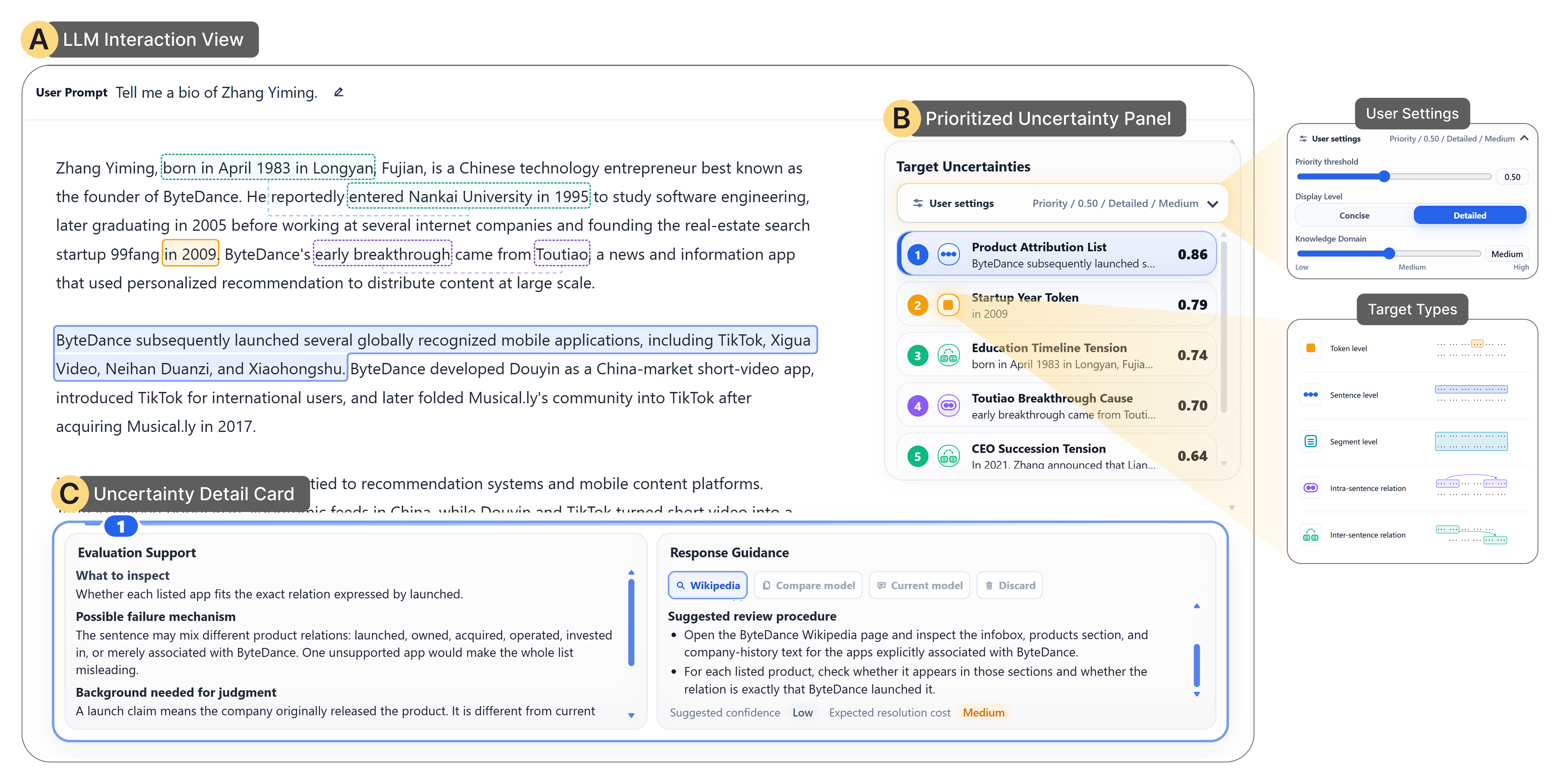}
\caption{U-Lens user interface. The main reading view overlays uncertainty targets in the generated response (DG1). The priority panel ranks and filters targets, with user settings for adjusting domain knowledge, display level, and ranking threshold (DG4). The detail card provides target-specific explanation and response guidance, with explanation depth adapted by the user settings (DG2-DG3).}
\label{figure:UI}
\end{figure*}

\subsection{User Interface}
As shown in \autoref{figure:UI}, U-Lens organizes uncertainty management as a reading-to-action workflow: users read the generated response, notice contextualized uncertainty targets, prioritize which targets deserve attention, inspect a selected target in more depth, and decide how to respond.

\subsubsection{LLM Interaction View (DG1)}
The \textit{LLM Interaction View} presents the user input together with the LLM-generated response, using the response as the primary reading surface and overlaying uncertainty targets directly onto the generated text. This instantiates DG1 by preserving the target's original context while supporting multiple textual structures identified in the formative study, including unit-based targets (token-level, sentence-level, and segment-level) and relation-based targets (relations within or between sentences). Selecting a highlighted target links the corresponding text span to the priority panel and opens its detail card, allowing users to move from contextual reading to focused evaluation.

\subsubsection{Prioritized Uncertainty Panel (DG4)}
The \textit{Prioritized Uncertainty Panel} instantiates DG4 by helping users allocate limited attention across multiple targets. Rather than requiring users to inspect highlighted spans in reading order, the panel presents targets in descending priority-score order. Each target card summarizes the concern with a short label, source snippet, target type, and priority cue. Users can adjust the priority threshold, scan lightweight summaries, and open the detail card when a target warrants closer inspection.

\paragraph{User Settings (DG2,DG4).}
At the top of the priority panel, U-Lens provides \textit{User Settings} for self-reported domain knowledge, display level, and priority threshold. These controls support DG4 by changing which targets are shown and how they are ordered, and support DG2 by changing the explanatory granularity shown for the selected target. For example, low domain knowledge foregrounds background context and term explanations, while higher domain knowledge foregrounds evidence, source grounding, and specific credibility concerns. Similarly, concise display settings show only the core concern and recommended action, while detailed settings reveal additional evidence needs, related context, and alternative interpretations.

\begin{figure*}[b]
\centering
\includegraphics[width=0.9\textwidth]{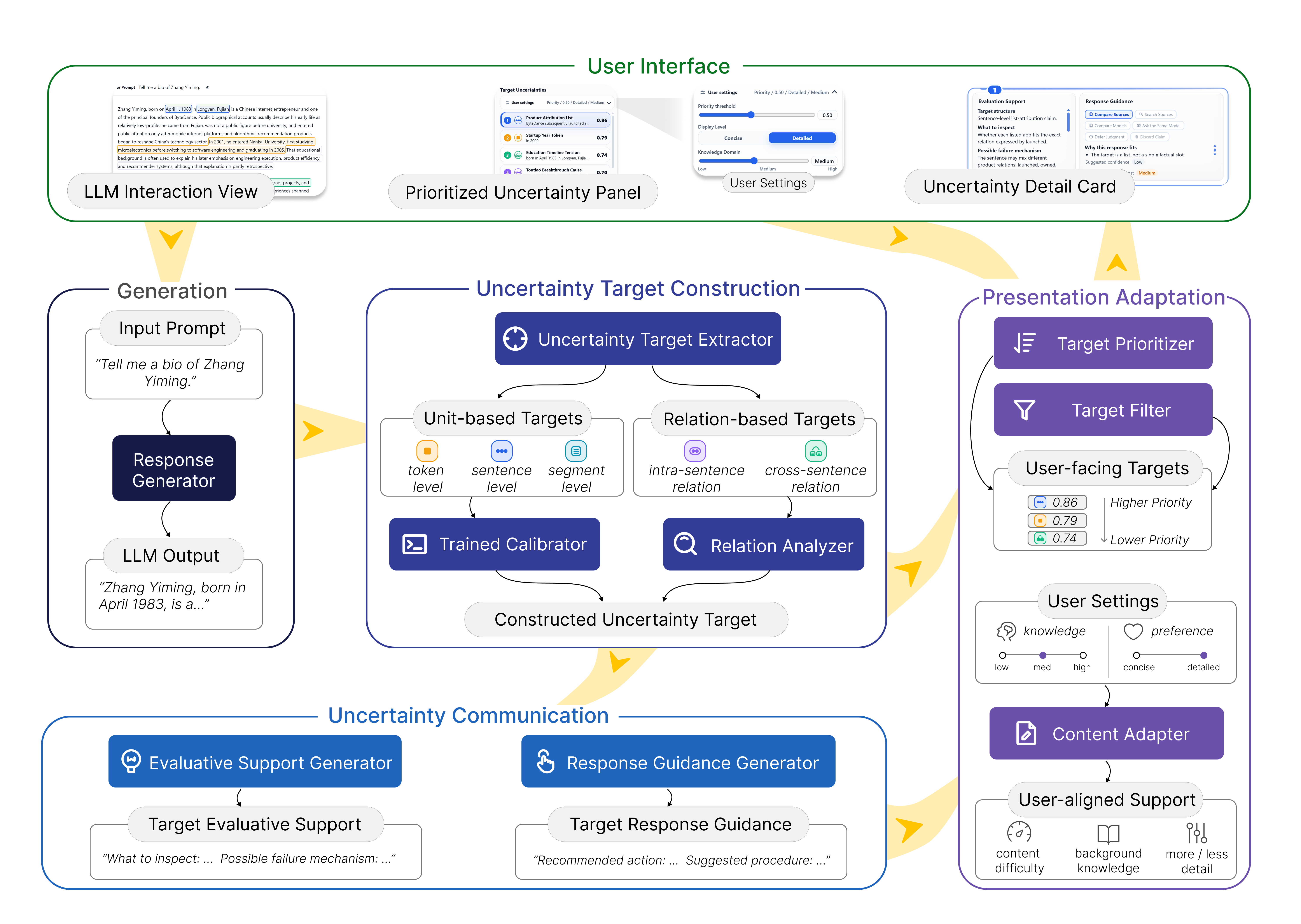}
\caption{U-Lens system architecture.}
\label{figure:system_architecture}
\end{figure*}

\subsubsection{Uncertainty Detail Card (DG2-DG3)}
The \textit{Uncertainty Detail Card} provides focused support for the selected target. Its evaluative explanation area instantiates DG2 by explaining what is uncertain, why the target may warrant scrutiny, what evidence or sources may help users judge it, and what related background knowledge may be useful. Its response guidance area instantiates DG3 by suggesting possible next steps, such as searching sources, comparing model responses, asking the same model for clarification, deferring judgment, or discarding the item, along with a short rationale and expected resolution cost or confidence when available.

\subsection{System Architecture}
\autoref{figure:system_architecture} shows an overview of the system architecture. The \textit{uncertainty target construction layer} constructs unit-based and relation-based uncertainty targets identified in the formative study (Section~\ref{section: formative study findings 1}). The \textit{uncertainty communication layer} transforms the extracted uncertainty targets into user-facing objects that include scores, labels, evaluation support, and response guidance. The \textit{presentation adaptation layer} uses user settings and target metadata to determine how uncertainty targets should be filtered, prioritized, and explained during interaction.

\begin{figure*}[b]
\centering
\includegraphics[width=0.9\textwidth]{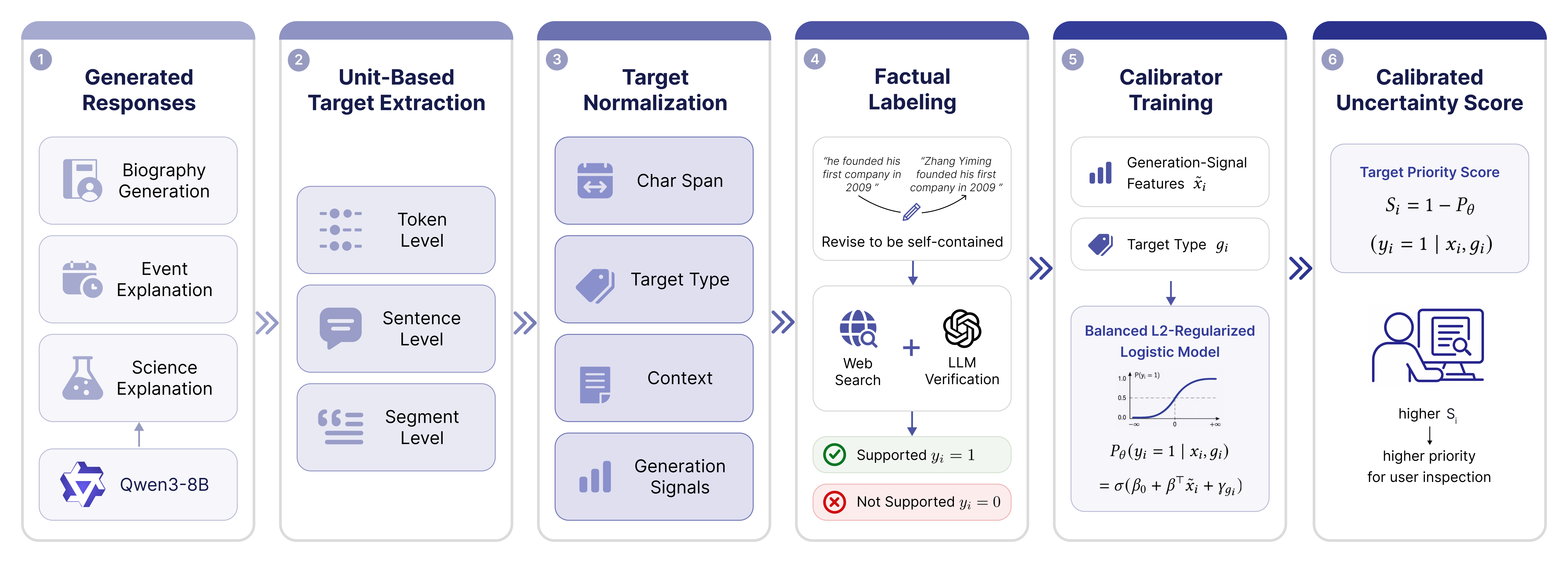}
\caption{Calibrator training for unit-based uncertainty targets.}
\label{figure:calibrator_training}
\end{figure*}

\subsubsection{Uncertainty Target Construction}
\label{section: Uncertainty Target Construction}
The uncertainty target construction layer converts the generated response into inspectable uncertainty targets. Rather than assuming that uncertainty is localized at a single textual unit, U-Lens extracts multiple types of targets identified in the formative study (Section~\ref{section: formative study findings 1}). These include \textit{unit-based uncertainty targets}, covering token-level, sentence-level, and segment-level texts, as well as \textit{relation-based uncertainty targets}, covering intra-sentence and inter-sentence relations. U-Lens uses grammatical rules and LLM-based extractors to identify candidate targets for each type and normalizes each candidate with its original character span, target type, surrounding context, and associated generation signals.

To score unit-based targets, we trained a generation-signal calibrator (\autoref{figure:calibrator_training}). The calibration corpus consisted of 7,059 target-level records derived from 150 LLM-generated responses across biography generation, event explanation, and scientific explanation tasks. These targets were labeled using web-search-assisted LLM judging, following existing long-form factuality evaluation pipelines \cite{min2023factscore, wei2024long}. We split the curated corpus at the response level: 120 generated responses were used for model selection and training, and 30 were used for model testing. At runtime, U-Lens applies the fitted calibrator using only generation-time signals and target type as inputs.

For target $i$, let $x_i$ denote the selected generation-signal vector and $g_i$ denote its target type. The calibrator is a class-balanced L2-regularized logistic model that estimates the probability that the target is supported:
\[
P_\theta(y_i=1 \mid x_i,g_i)
= \sigma(\beta_0+\beta^\top \tilde{x}_i+\gamma_{g_i}),
\]
where $y_i=1$ denotes a supported target, $\tilde{x}_i$ is the standardized feature vector, and $\gamma_{g_i}$ is the target-type term. 

U-Lens uses the complement as the target priority score:
\[
S_i = 1 - P_\theta(y_i=1 \mid x_i,g_i).
\]
Higher $S_i$ values indicate lower estimated factual support and therefore higher priority for user inspection. U-Lens uses the score to support user-facing prioritization rather than automated factuality judgment. Details of model selection and calibrator implementation are provided in Appendix~\ref{appendix:calibrator-details}.

For relation-based targets, relational uncertainty is defined as a potential text-internal tension that may require reconciliation, so each accepted target must align with a contiguous span in the original response and contain at least two verbatim textual pieces that constitute the tension. U-Lens uses the extractor's confidence that the span is a plausible relation target as a priority cue, without treating it as evidence that the content is factually incorrect.

\subsubsection{Uncertainty Communication}
The uncertainty communication layer generates user-facing information for each uncertainty target. For each target, U-Lens first constructs a compact evidence basis from the target construction output, including its textual grounding, priority evidence, and type-specific rationale. For unit-based targets, this evidence is grounded in calibrated generation signals; for relation-based targets, it is grounded in the extracted relation components and their rationale.

U-Lens then converts this evidence basis into structured communication objects shown in the \textit{Uncertainty Detail Card}, including evaluative support and response guidance. The evaluative support organizes the evidence basis around what should be inspected, why the target was surfaced, what background or source context may be useful, and what alternative interpretations may remain possible. For response guidance, U-Lens first defines an action selection knowledge base derived from the formative study, such as using web search for directly checkable factual details and follow-up questions for ambiguous textual spans. U-Lens generates the selected action with its rationale, suggested review procedure, suggestion confidence, and expected resolution cost.

\subsubsection{Presentation Adaptation}
The presentation adaptation layer applies two forms of session-level adjustment. Target-level adaptation determines which uncertainty targets are shown and in what order. Content-level adaptation determines which parts of each target's communication object are shown and how they are phrased for the user's topic familiarity.

For target-level adaptation, U-Lens uses the target priority score and the priority threshold selected in the user settings. Targets are ordered by descending priority score, and the threshold filters out targets below the current inspection level. Together, this score and threshold produce the visible target list shown in the \textit{Prioritized Uncertainty Panel}.

For content-level adaptation, U-Lens renders different views of the same communication object. Display level selects the visible fields: the concise view keeps the target label, core concern, recommended action, and resolution cost, while the detailed view adds rationale, background needs, evidence boundaries, and alternative interpretations. Topic familiarity selects a phrasing profile. Lower-familiarity views use fewer specialized terms and add brief clarifications; higher-familiarity views preserve more domain-specific terms and foreground evidence needs, source-grounding cues, failure mechanisms, and unresolved credibility concerns. U-Lens realizes these profiles through a constrained LLM rewrite over the selected fields, with instructions not to add factual claims or alter the target, priority evidence, or recommended action.

\subsection{Implementation}
We implemented U-Lens as a web-based prototype with a Python backend and a React frontend built with TypeScript and Vite. The backend generates long-form LLM responses, records generation-time signals, constructs uncertainty targets, and produces user-facing uncertainty support. Because U-Lens relies on model-internal signals, we use Qwen3-8B as the base generation model. Offline weak labels for calibrator training are produced using a web-search-enabled GPT-5.4 judge. User-facing content adaptation is limited to constrained GPT-5.4-mini rewrites of existing support fields, with no web search.
For evaluation, we pre-generated study materials with the backend to select comparable materials and ensure consistency across conditions, while preserving the same generation and uncertainty-construction pipeline used in the prototype.

%% file: data/chap5.tex
\section{User Evaluation}
We evaluated U-Lens through a controlled within-subjects study that examined whether it better supported user uncertainty management during limited-budget factual verification than a confidence-cue baseline. Rather than asking participants to exhaustively verify every claim in a long-form LLM response, we placed them in a realistic constrained-review scenario: they had to decide which statements most deserved attention, spend a limited number of verification opportunities, and prepare a more reliable handoff artifact for another person to use.

\subsection{Participants}
We recruited 18 participants who reported regular use of LLM systems for knowledge-intensive tasks such as learning unfamiliar topics, preparing background materials, or supporting writing and research activities. Participants included 10 males and 8 females, aged 18-26 years ($M=22.22$, $SD=2.16$). Each participant received \$20 upon completing the study. Detailed demographic and background information is provided in \autoref{tab:user_evaluation_demographic}.   

\begin{table*}[]
\caption{Demographic and background information for participants in the user evaluation. LLM Skepticism and Material Familiarity were measured on 7-point scales; material familiarity shows self-reported average familiarity across the two assigned materials.}
\label{tab:user_evaluation_demographic}
\small
\begin{tabular}{ccccccc}
\toprule
\textbf{ID} & \textbf{Gender} & \textbf{Age} & \textbf{Education} & \textbf{Verification Habit} & \textbf{LLM Skepticism} & \textbf{Material Familiarity} \\
\midrule
P1 & Male   & 24 & PhD        & Rarely    & 5 & 2.0 \\
P2 & Female & 24 & Bachelor's & Rarely    & 4 & 3.0 \\
P3 & Female & 25 & PhD        & Sometimes & 5 & 2.5 \\
P4 & Female & 22 & Bachelor's & Often     & 6 & 1.5 \\
P5 & Male   & 20 & Bachelor's & Rarely    & 3 & 2.0 \\
P6 & Female & 19 & PhD        & Sometimes & 5 & 4.0 \\
P7 & Male   & 23 & Master's   & Sometimes & 4 & 3.0 \\
P8 & Female & 21 & Bachelor's & Sometimes & 5 & 2.5 \\
P9 & Female & 23 & Master's   & Sometimes & 4 & 2.0 \\
P10 & Male   & 18 & Bachelor's & Sometimes & 5 & 1.0 \\
P11 & Female & 19 & Bachelor's & Sometimes & 5 & 1.5 \\
P12 & Female & 22 & Bachelor's & Often     & 6 & 2.0 \\
P13 & Male   & 23 & Master's   & Often     & 4 & 2.5 \\
P14 & Male   & 22 & Bachelor's & Sometimes & 5 & 3.0 \\
P15 & Male   & 23 & Master's   & Often     & 6 & 2.0 \\
P16 & Male   & 24 & Master's   & Sometimes & 5 & 2.5 \\
P17 & Male   & 26 & PhD        & Sometimes & 4 & 3.0 \\
P18 & Male   & 22 & Bachelor's & Sometimes & 5 & 1.0 \\
\bottomrule
\end{tabular}
\end{table*}

\subsection{Study Materials and User Task}
We prepared 12 long-form LLM-generated responses across the same three task types used in the formative study: biography generation, event explanation, and scientific explanation. The materials were distinct from the formative-study examples. To build a controlled material pool, we generated five responses for each of 30 candidate task prompts and screened the resulting 150 responses based on length, sentence count, readability, and annotated factual-issue count. Factual-issue annotations were generated using existing long-form factuality evaluation pipelines \cite{min2023factscore, wei2024long}, then cross-checked and supplemented by GPT-5.4, and manually verified by one HCI researcher. This evaluation pool was constructed separately from the 150-response calibration corpus (Section~\ref{section: Uncertainty Target Construction}), with no overlap in task prompts. We then selected the final materials so that responses within each task type were comparable and avoided trials that were either nearly flawless or overwhelmingly inaccurate. 

The main user task was framed as a handoff scenario. Participants were told to imagine that they were preparing background material for a group assignment and had to verify an LLM-generated response before passing it to a teammate. Because time and verification opportunities were limited, they could not check everything. In each trial, participants selected the statements they considered most important to verify, spent their limited verification budget on those statements, and prepared a final handoff version with brief notes about revisions or unresolved concerns.

\subsection{Conditions}
\begin{figure*}[htb]
\centering
\includegraphics[width=1.0\textwidth]{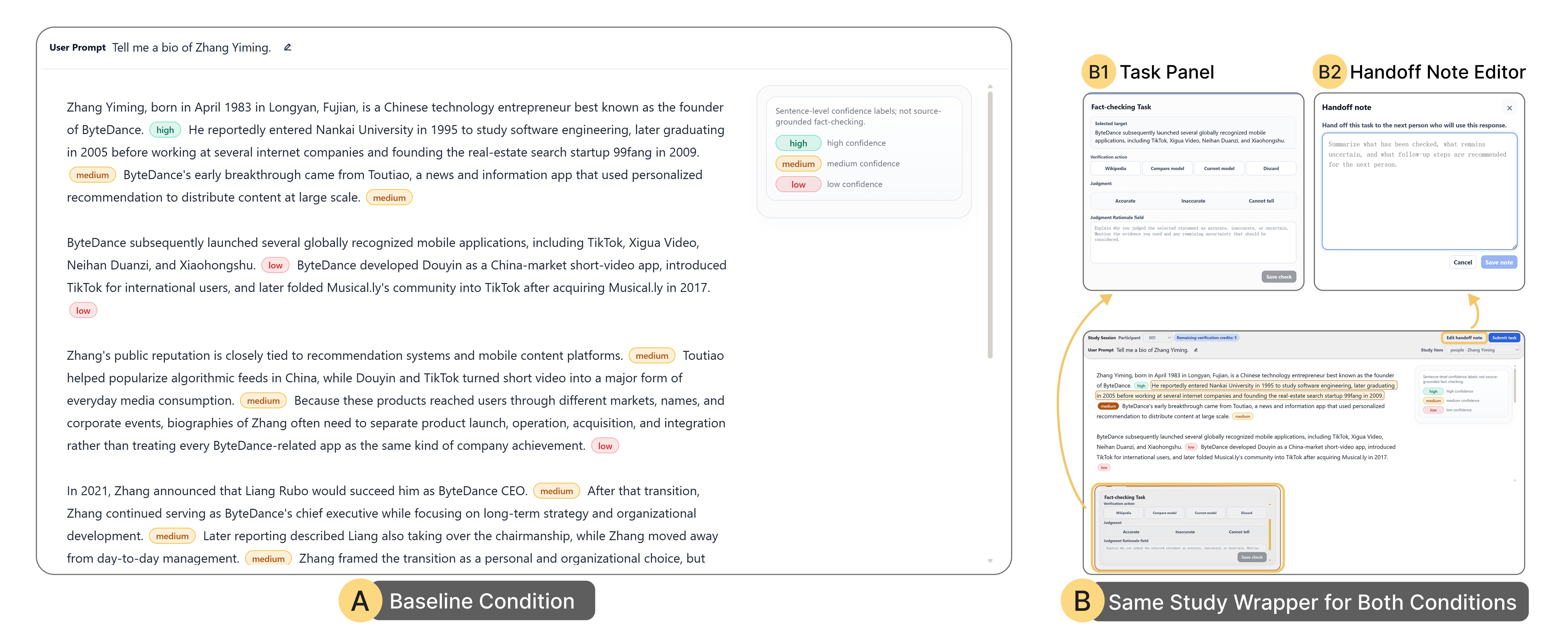}
\caption{Evaluation study interfaces. (A) The confidence-cue baseline displays lightweight low, medium, or high estimated-support labels for sentences. (B) Both U-Lens and baseline conditions are embedded in the same study wrapper.}
\label{figure:user_evaluation_UI}
\end{figure*}

We compared two conditions in a within-subjects design.
We selected the baseline to represent a cue-centered comparator rather than a no-support control. This choice was closest to prior systems that present factuality scores or visual factuality indicators for generated content \cite{do2024facilitating, do2025highlight}, as well as work that uses ordinal confidence labels to support hallucination inspection \cite{leiser2024hill}. Following work on verbalized and qualitative uncertainty communication \cite{dhami2022communicating, steyvers2025large, padilla2021uncertain}, the baseline showed categorical sentence-level support labels.

\textbf{Confidence-cue baseline.} 
The baseline presented the same LLM-generated response as U-Lens, but only showed one estimated-support label for each sentence (see \autoref{figure:user_evaluation_UI}). For a fair comparison, it used the same generated output and the same generation-signal calibrator used for U-Lens unit-based targets. For each sentence $i$, the calibrator estimated the supported probability $c_i = P_\theta(y_i=1 \mid x_i, g_i=\text{sentence-level})$. The interface mapped this internal score into the three labels: low, medium, or high support. This baseline represented a minimal confidence-cue interface without fine-grained uncertainty-management support.

\textbf{U-Lens.}
The U-Lens condition presented the full interaction design introduced in Section \ref{section: DESIGN AND IMPLEMENTATION}. It provided contextual uncertainty targets, the prioritized target panel, and target-specific detail cards with evaluative support and response guidance.

For the evaluation, both condition-specific interfaces were embedded in the same study wrapper. The wrapper displayed the remaining action budget, provided the same controlled actions, collected participants' response decisions and brief rationales for selected statements, and supported final handoff submission (\autoref{figure:user_evaluation_UI}). The wrapper provided the same controlled actions in both conditions: opening the assigned Wikipedia page for the material, asking the comparison model, asking the original generation model with the task context, deferring judgment, or using ``discard'' when participants could directly judge an item from their own knowledge without using an additional resource. Participants could record only one action for each selected item. These wrapper controls were used only to support task completion and data collection. They were held constant across conditions and were not part of the experimental manipulation.

\subsection{Study Procedure}

\textbf{Introduction.}
Participants first provided informed consent and completed the pre-study questionnaire. The facilitator then introduced the handoff scenario, the verification task, and the two interface conditions.

\textbf{Main user task.}
Participants completed two formal trials: one with U-Lens and one with the confidence-cue baseline. Each participant was randomly assigned to one task category and viewed two matched responses on different topics within that category, one for each condition. Condition order and material–condition assignment were counterbalanced across participants.

In each trial, participants read the prompt and LLM-generated response, selected content they considered important to verify, chose one controlled action for each selected item, and recorded a judgment with a rationale and any correction or handoff note. They had to spend five verification credits before submitting the trial. After the five records were saved, they wrote a brief handoff note summarizing what they checked and what remained uncertain.

\textbf{Post-condition questionnaire.}
After each trial, participants completed a post-condition questionnaire about the just-completed condition. It included 7-point Likert-scale NASA-TLX items, items about task outcome and confidence, and uncertainty-management statements covering interpretation, evaluation, decision, and cross-stage support.

\textbf{Semi-structured interview.}
At the end, we conducted a comparative semi-structured interview. The interview asked how participants identified content worth inspecting, how they used the interface cues and detail-card information, how they chose verification actions, and how they allocated the limited verification budget. We also asked about workload, judgment agency, information overload, and what they would keep or change for real limited-budget verification work.

\begin{figure*}[htb]
\centering
\includegraphics[width=0.9\textwidth]{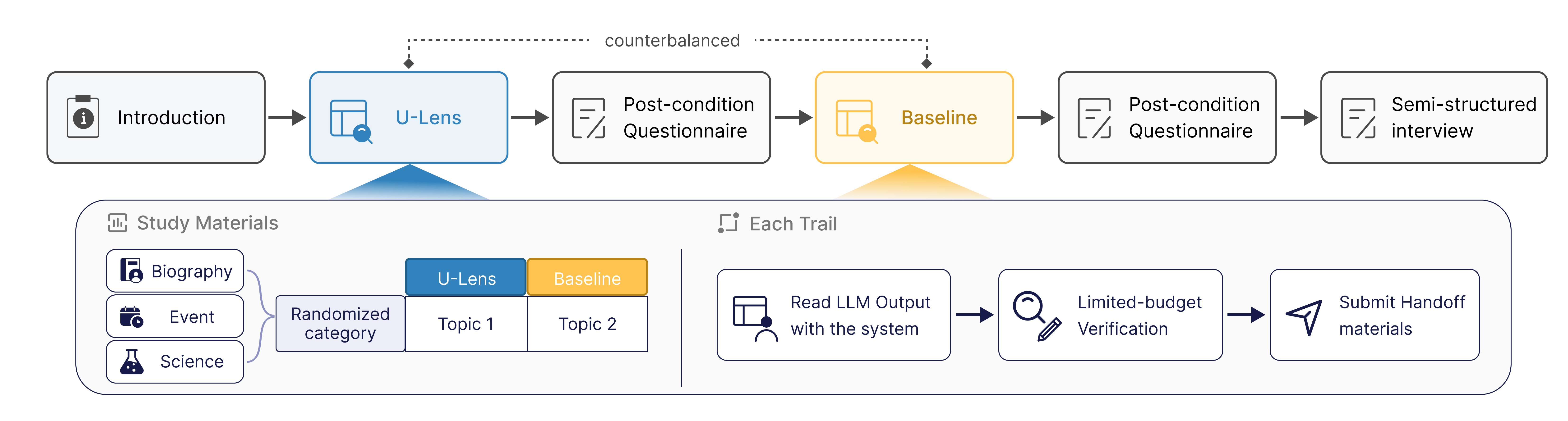}
\caption{Evaluation study procedure.}
\label{figure:evaluation_study_procedure}
\end{figure*}

%% file: data/chap6.tex
\section{Results}

\subsection{Overall Task Outcomes}

\subsubsection{Objective Task Outcomes}

\begin{figure*}[htbp]
\centering
\includegraphics[width=1.00\textwidth]{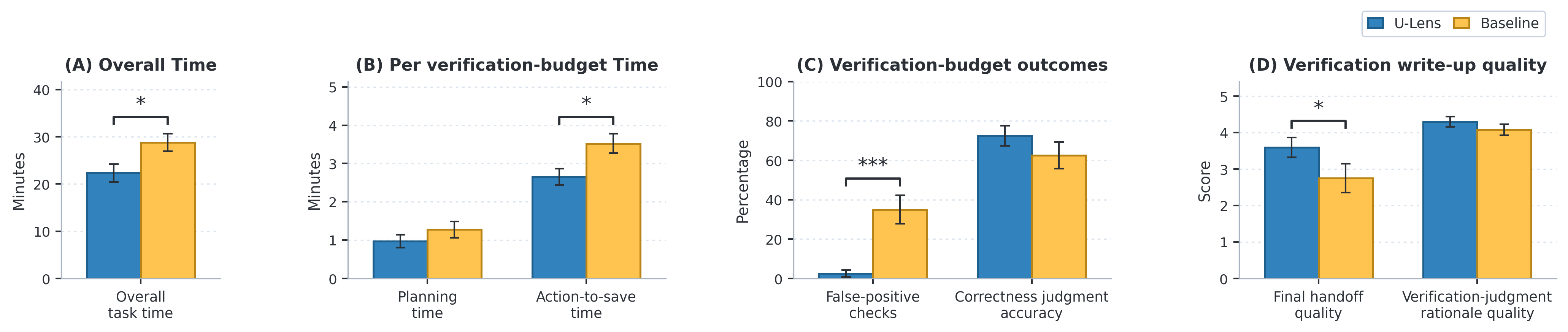}
\caption{Objective task outcomes by condition. (A) Overall task time. (B) Time spent within each saved verification-budget check, including action-planning time and action-to-save time. (C) Verification-budget outcomes, including the percentage of the five-check budget spent on false-positive checks and final correctness-judgment accuracy. (D) Quality of participant-entered verification text fields: the final condition-level handoff note and the per-check verification-judgment rationale fields. Error bars show standard error. Asterisks indicate paired-test significance (* $p < .05$, ** $p < .01$, *** $p < .001$).}
\label{fig:evaluation-objective-barplots-combined}
\end{figure*}

Participants completed the overall task in less time with U-Lens ($M=22.32$ minutes, $SD=7.56$) than with the confidence-cue baseline ($M=28.81$ minutes, $SD=7.38$), $p<.05$ (\autoref{fig:evaluation-objective-barplots-combined}A). Within each verification workflow, the average time from selecting an action to saving the resulting decision was also lower with U-Lens ($M=2.65$ minutes, $SD=0.86$) than with the baseline ($M=3.52$ minutes, $SD=1.02$), $p<.05$. The action-planning time ($M=0.97$, $SD=0.68$ vs. $M=1.27$, $SD=0.84$) showed no significant difference (\autoref{fig:evaluation-objective-barplots-combined}B).

For the five verification-budget checks completed by each participant, U-Lens led to fewer false-positive checks ($M=0.13$, $SD=0.34$)---defined as verification attempts spent on targets that did not contain actual errors---compared with the baseline ($M=1.75$, $SD=1.44$), $p<.001$. Other objective outcomes did not differ significantly between conditions (\autoref{fig:evaluation-objective-barplots-combined}C). Per-check correctness judgment accuracy was numerically higher with U-Lens ($M=72.50\%$, $SD=20.49\%$) than with the baseline ($M=62.50\%$, $SD=27.20\%$), but the difference was not significant. The number of distinct action types used across the five checks was also similar between U-Lens ($M=1.81$, $SD=0.66$) and the baseline ($M=1.75$, $SD=0.58$).

We also coded two participant-entered text fields: (1) the final condition-level handoff note, which captured how participants summarized what had been checked or modified, articulated remaining uncertainty, and provided concrete follow-up work or verification methods; and (2) the per-check verification-judgment rationale fields,  which captured the written reasoning quality underlying each saved verification judgment. Each field was rated on a clearly defined 0--5 rubric by two HCI researchers independently. The two coders showed high agreement on both fields ($ICC(2,1)=0.94$ for handoff notes; $ICC(2,1)=0.93$ for verification-judgment rationales). We used the average of the two coders' ratings in subsequent analyses. As shown in \autoref{fig:evaluation-objective-barplots-combined}D, U-Lens produced higher final handoff quality scores ($M=3.59$, $SD=1.07$) than the baseline ($M=2.75$, $SD=1.57$), $p<.05$. Verification-judgment rationale quality showed no significant difference between U-Lens ($M=4.29$, $SD=0.56$) and the baseline ($M=4.08$, $SD=0.60$). These results suggest that U-Lens primarily helped participants organize and communicate the overall state of uncertainty verification at handoff, rather than substantially improving how they reasoned about each judgment’s correctness using external evidence.

\subsubsection{Subjective Task Outcomes}
\begin{figure*}[htbp]
\centering
\includegraphics[width=0.65\textwidth]{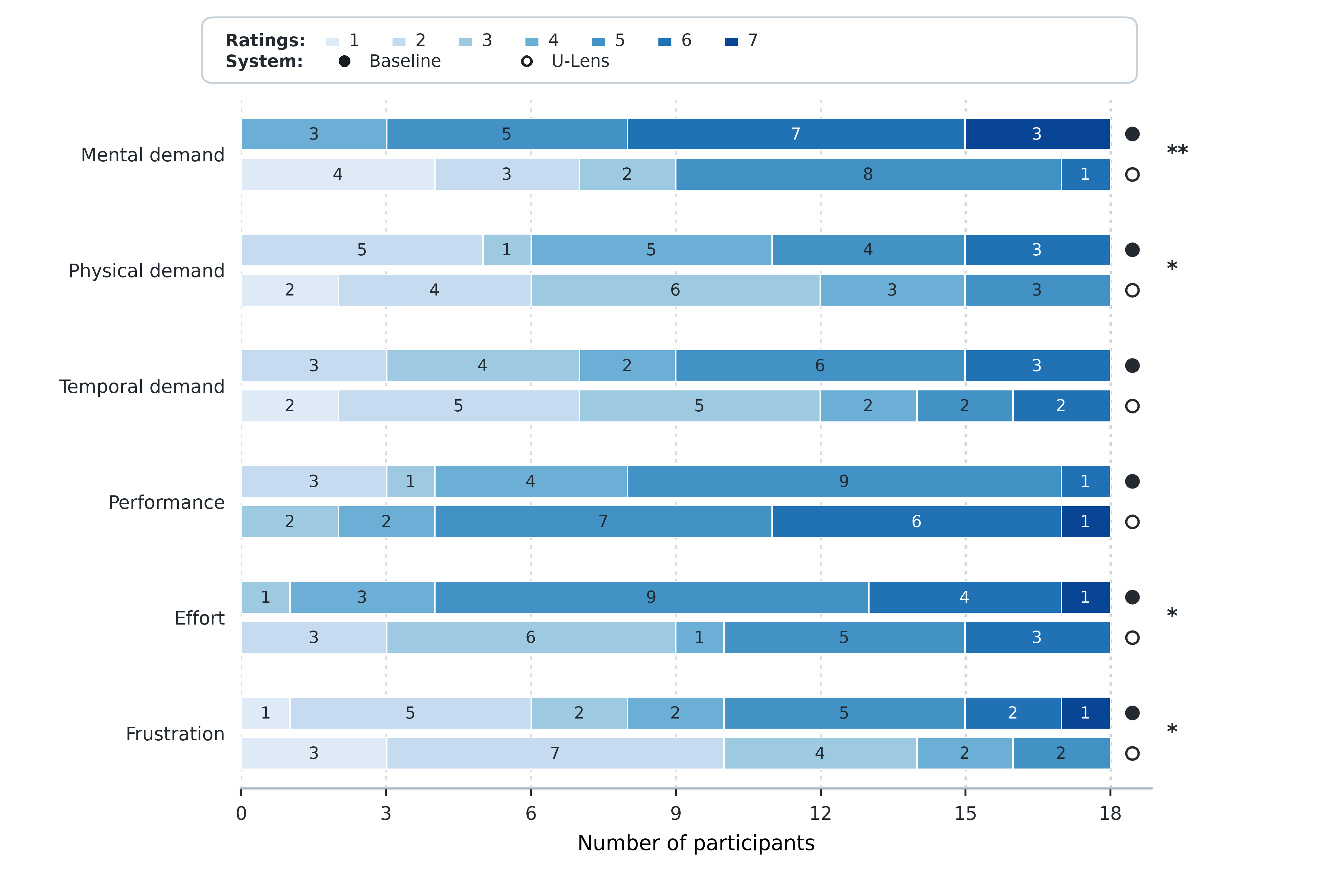}
\caption{NASA-TLX workload ratings by condition. Compared with the baseline, U-Lens shifted participants' ratings toward lower workload, with significant reductions in mental demand, physical demand, effort, and frustration. Asterisks indicate statistical significance (* $p < .05$, ** $p < .01$, *** $p < .001$).}
\label{fig:evaluation-nasa}
\end{figure*}

We analyzed NASA-TLX dimensions using Wilcoxon signed-rank tests with Holm correction. Participants reported significantly lower workload with U-Lens than with the baseline on mental demand ($M=3.94$, $SD=1.35$ vs. $M=5.56$, $SD=0.98$), $p<.01$; physical demand ($M=3.06$, $SD=1.26$ vs. $M=3.94$, $SD=1.47$), $p<.05$; effort ($M=3.94$, $SD=1.43$ vs. $M=5.06$, $SD=0.94$), $p<.05$; and frustration ($M=2.61$, $SD=1.24$ vs. $M=3.83$, $SD=1.76$), $p<.05$. No significant differences were observed for temporal demand ($M=3.17$, $SD=1.54$ vs. $M=4.11$, $SD=1.41$) or perceived performance ($M=5.11$, $SD=1.08$ vs. $M=4.22$, $SD=1.22$).

Interview results further suggested that U-Lens reduced the overall task workload. 15/18 participants explicitly said that U-Lens reduced their workload. Participants no longer had to decide from scratch which statements deserved the limited verification credits. P12 explained that U-Lens removed the need to perform ``the most basic priority ranking'' over sentences, and P3 said that, with the baseline, they could not ``immediately find five candidates by intuition'' and needed more time for ``thinking and locating.'' P5 said they could avoid first reading the whole response and instead ``directly check this point'' because the target was clear. The lower effort followed from this clearer entry point into action: P18 described U-Lens as helping them ``easily find what I should check'' and effectively skipping the step of ranking candidate checks.

\subsection{How U-Lens Supported the Three Stages of the Uncertainty Management Process}
\begin{figure*}[htbp]
\centering
\includegraphics[width=0.65\textwidth]{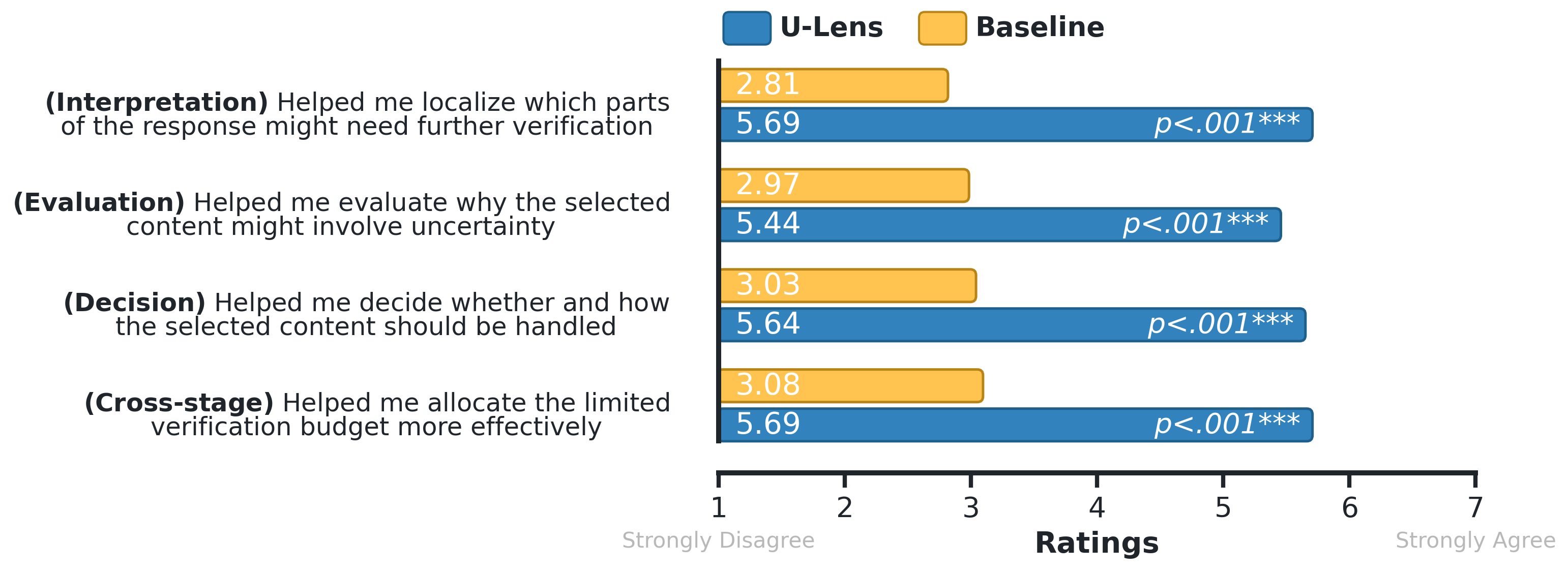}
\caption{Perceived support for the uncertainty-management process. Participants rated U-Lens as more helpful than the confidence-cue baseline for localizing uncertain content, evaluating why it might be uncertain, deciding how to handle it, and allocating the limited verification budget across stages.}
\label{fig:evaluation-stage-support}
\end{figure*}

Questionnaire ratings showed that participants perceived U-Lens as providing stronger support than the baseline across the overall uncertainty-management process ($M=5.69$, $SD=1.00$ vs. $M=3.08$, $SD=1.60$) and all three TMIM stages: interpretation ($M=5.69$, $SD=0.73$ vs. $M=2.81$, $SD=1.46$), evaluation ($M=5.44$, $SD=0.89$ vs. $M=2.97$, $SD=1.56$), and decision ($M=5.64$, $SD=1.00$ vs. $M=3.03$, $SD=1.71$). Wilcoxon signed-rank tests showed that all differences were significant ($p<.001$; \autoref{fig:evaluation-stage-support}).

In this section, we further analyze interaction logs and interview data to examine how U-Lens supported participants across each stage of the uncertainty management process---interpretation, evaluation, and decision.

\subsubsection{Interpretation Stage: U-Lens Reduced Open-ended Inspection by Prioritizing and Specifying Uncertainty Targets}
In the interpretation stage, U-Lens helped participants reduce the openness of initial inspection by turning long-form responses into fine-grained, prioritized uncertainty targets.

U-Lens did not increase the amount of inspection: the number of unique inspected targets was comparable between U-Lens ($M=10.44$, $SD=2.34$) and the baseline ($M=11.88$, $SD=9.73$). Instead, U-Lens changed the quality and order of inspection. Inspection precision was significantly higher with U-Lens ($M=92.41\%$, $SD=8.74\%$) than with the baseline ($M=50.23\%$, $SD=23.34\%$), $p<.001$. Within U-Lens trials, inspected targets had an average priority rank of $M=6.00$ ($SD=0.98$), and the first inspected target had an average rank of $M=5.44$ ($SD=4.86$). These results suggest that U-Lens helped participants follow a more focused, priority-guided inspection path, rather than inspecting the response sentence by sentence.

Interview data further explained how participants formed this inspection path. Most participants described using U-Lens' priority list, scores, and highlights to form an initial inspection path. 16/18 participants described using these cues to decide where to look first, often after a brief skim of the response. For example, P8 noted that the priority list narrowed the inspection space: ``The text itself was very long... but the points listed on the right were only 12... so I could roughly know where the biggest questions were.'' Several participants also combined this priority-guided path with their own knowledge. As P17 explained, ``First, I relied on what I already knew... second, I followed the items that the model gave higher priority to.''  By contrast, participants noted that the baseline’s sentence-level confidence cues often marked too many sentences, leaving them to determine the inspection path on their own. Therefore, U-Lens reduced the openness and burden of locating potential problems in long-form LLM responses.

U-Lens also helped participants move from vague, coarse-grained suspicion to smaller inspectable targets. 9 participants reported that U-Lens made the potential problem within a sentence more concrete (P3: ``it broke a sentence down more finely and directly told me which part of the sentence needed closer checking.'') This helped participants ``focus more closely on a specific part'' (P18) and reduced the need to infer ``which part of the sentence was wrong'' (P9), allowing them to enter the subsequent evaluation stage with a clearer object of judgment.

In addition, U-Lens expanded what participants considered uncertain. Thirteen participants said that U-Lens surfaced factual details they would normally assume to be correct. For example, P8 noted that details such as birthplace and childhood location were easy to take for granted: ``if the system had not marked them, I might have assumed that these details would not be problematic.'' Some participants also noted that U-Lens drew attention to conceptual or relational issues beyond isolated facts. For instance, P14 described one target as concerning ``the relation between events, or the logic'', adding that the sentence ``read quite smoothly, but I had not carefully considered it from that angle.''

However, 6 participants raised concerns that focusing on system-suggested targets could fragment their reading and reduce their holistic understanding of the LLM response. Targeted reading could also overly narrow users’ attention, leading them to overlook unmarked content. For example, P11 reflected, ``my understanding of it became rather fragmented... I did not read the unmarked content very carefully''. This suggests a tension between making inspection more focused and preserving users’ broader engagement with the full response.

\subsubsection{Evaluation Stage: U-Lens Turned Marked Targets into Checkable Error Hypotheses}
In the evaluation stage, U-Lens helped participants turn a marked target into a more checkable error hypothesis. Rather than only indicating that a span was uncertain, the \textit{Uncertainty Detail Card} described possible ways in which the target might be wrong. 17/18 participants said it helped them identify what was at issue in a selected target, and 9/18 said it changed or increased their suspicion. For example, P10 noted, ``I might have looked at this sentence before and thought there was no problem, but after it provided some support, I felt that there might still be something problematic.'' P1 similarly valued explanations such as ``the location might have been confused,'' which was enough to make the target worth checking: ``When I saw this sentence, I knew that this might need to be checked.''

U-Lens also helped participants see that one marked target could contain multiple hidden sub-claims. This was especially useful for dense sentences or relation-heavy statements. 5 participants described how the evaluation support decomposed a target into multiple elements or clarified their relations. For example, P8 described a target about legal work as containing several separate points: ``San Francisco as the location, the law firm as the workplace, and how many firms there were. Was it one firm or several? Each point indeed needed to be checked separately.'' This helped participants evaluate the internal structure of a target rather than treating it as a single factual claim.

The value of evaluation support also depended on users’ knowledge and the complexity of the material. 4 participants said they needed more explanation for unfamiliar domains or reasoning-heavy content, whereas simple factual targets often required less support. P18 explained that, for an unfamiliar astronomy topic, they needed U-Lens because ``I did not know which things were actually worth checking''. By contrast, P10 noted that the baseline could sometimes be sufficient for basic factual details such as dates or people, whereas U-Lens was more useful for ``causal relationships and deeper-level reasons'', especially when the uncertainty concerned intra- or inter-sentence relations.

However, participants did not find all explanation fields equally useful. 8/18 participants commented that the explanations contained too much information or that some fields overlapped. For example, P3 said, ``The most critical thing, and what I care about most, is why the system judged it to be problematic. It does not need to be as detailed as it is now.'' This suggests a tension between providing enough explanation for difficult targets and allowing users to control how much detail they receive.

\subsubsection{Decision Stage: U-Lens Scaffolded Verification Decisions Without Replacing User Judgment}
In the decision stage, U-Lens helped participants move from a selected verification action to a saved judgment more efficiently, without narrowing the set of actions they used. The average action-to-save time was significantly lower with U-Lens ($M=159.28$ seconds, $SD=51.48$) than with the baseline ($M=211.45$ seconds, $SD=61.35$), $p<.05$. This efficiency gain was especially visible in Wikipedia checks, the most frequently used action---the time from opening Wikipedia to saving the check was lower with U-Lens ($M=178.40$ seconds, $SD=65.09$) than with the baseline ($M=246.93$ seconds, $SD=94.17$), $p<.05$. At the same time, the number of distinct action categories used across the five saved checks was similar between U-Lens ($M=1.81$, $SD=0.66$) and the baseline ($M=1.75$, $SD=0.58$). Together, these results suggest that U-Lens helped participants follow a more efficient verification path, rather than pushing them toward a narrower set of actions.

Logs and interviews further show that this support did not replace participants' own judgment. Across 90 saved U-Lens checks, participants chose the primary recommended action in 43 checks (47.8\%). Interviews also suggested that participants varied substantially in how they adopted the response guidance. Some relied mainly on their own tool preferences. For example, P16 explained, ``If this part is more factual, I would be more inclined to use Wikipedia,'' and avoided asking the current model because it might ``force an explanation that it is correct''. Others directly followed the recommended action. As P7 explained, ``After there was recommendation information... I would directly follow its recommended action and do that''. A third group treated the recommendation as a candidate plan and judged whether it fit the target. As P8 put it, ``I did not need to think from scratch about what to do. I only needed to decide whether what it said was reasonable; if it was reasonable, I could just follow it''.

Even when U-Lens did not change participants' action choices, it still made verification more actionable by helping them formulate queries (P9: ``It did not affect my choice very much... but it affected the way I asked the question.'') 10/18 participants said the system helped them decide how to ask a model or what relation to look for when checking webpages. For example, P17 said, ``It prompted me to first ask what institutions they each worked at, and whether there was any publicly verifiable relationship between them.'' Thus, response guidance was useful not only as a tool recommendation, but also as a way to translate an uncertainty appraisal into a concrete verification question.

After verification, participants made final judgments based on the state of evidence rather than the system recommendation itself. 14/18 participants described evidence-based rules for deciding whether a target was accurate, inaccurate, or unverifiable. Participants also considered source reliability. For example, P8 said, ``If the information was verified on Wikipedia, then when there was a conflict, I would by default treat Wikipedia as correct.'' In contrast, model-based evidence was treated more cautiously (P18: ``Even if it expressed the same thing twice, I would still think this was impossible to judge.'') These accounts show that U-Lens supported the verification decision process, while participants retained control over the final correctness judgment.

\subsection{How Users Positioned U-Lens as an Uncertainty Management Tool}

\subsubsection{U-Lens Reduced Verification-Organization Burden, but Added Information and Oversight Burden}
Participants' overall reflections suggested that U-Lens was not experienced as simply reducing or increasing work. Rather, it redistributed the burden involved in managing uncertainty.

17/18 participants described U-Lens as reducing the organizational burden of verification. Instead of first deciding which details in a long response might deserve checking, participants could begin from a smaller set of system-suggested targets and ``directly check this starting point'' (P5). P11 said that they no longer needed to spend as much ``thinking and effort'' on deciding whether every detail needed verification. P14 similarly noted that ``the workload was not substantially reduced, but the amount of thinking was reduced somewhat''.

However, this organizational support also introduced additional information to manage. Several participants said they selectively ignored parts of the system output when it felt unnecessary. As P3 put it, ``if I feel there is too much, I just will not read it.'' For others, reading the added support text itself became a cost. P18 explained that the detail card included both support and guidance, and that ``there were quite a lot of words'', so reading them also took time.

Some participants also described an oversight burden. Because U-Lens narrowed the inspection targets, they sometimes felt responsible for checking whether the system had excluded something important. P15 said that when they were not in a hurry, U-Lens could increase the work because they would spend extra effort checking ``whether there were issues I cared about in the parts that were not boxed''.

\subsubsection{Users Valued U-Lens for Focused Checking, with Concerns About Autonomy and Coverage Boundaries}
Participants valued U-Lens most as a tool for focused checking. P8 described this as reducing ``dozens or hundreds of sentences'' to ``a dozen or so core points of doubt''. P5 also valued that U-Lens did not merely mark low confidence, but made the object of doubt clearer: ``You say this sentence is low-confidence, but what exactly are you questioning?'' These accounts position U-Lens as a practical entry point into focused uncertainty verification.

At the same time, participants noted autonomy concerns in this focused checking mode. The concern was not only whether U-Lens made the final judgment, but whether it pre-structured what users would think about in the first place. P16 said that part of their cognition ``had been handed over'', and that they might stop reading the full text and only focus on marked problematic sentences, making their sense of autonomy weaker. P10 similarly noted that U-Lens made the process easier because much of the review was automated, but this also meant that ``it did not require too much thinking on my own''.

Participants also described concerns about coverage boundaries. The same narrowing that made U-Lens efficient could also lead users to ``ignore other places that might have errors'' (P12). P18 noted that if a factual error appeared in a sentence U-Lens did not mark, ``there is no way I would verify that sentence''. P11 noted that U-Lens was helpful if the goal was fact-checking an LLM response, but if the task required deeper understanding of the content, their ``overall grasp would be much worse''. Thus, U-Lens helped manage uncertainty about factual correctness rather than difficulties in deeper content understanding.

%% file: data/chap7.tex
\section{Discussion}
Across HCI domains, users increasingly work with AI systems in situations where they must act under uncertainty, including writing \cite{lee2022coauthor, li2024value}, learning \cite{jin2024teach}, decision support \cite{sivaraman2023ignore, yang2023harnessing}, code generation \cite{mowar2025codea11y}, and LLM response verification \cite{cheng2024relic, leiser2024hill}. Prior work has supported users through confidence indicators, explanations, inconsistency cues, source attributions, and citations \cite{leiser2024hill, kim2025fostering, cheng2024relic, do2024facilitating, gao2023enabling}. These approaches make potential problems more visible and generated answers better supported, but much of this support still centers on the answer itself. We extend this line of work by focusing on how users make sense of and act on uncertain AI-generated information.

In long-form LLM responses, users need not only to notice uncertain content, but also to determine what the uncertainty refers to, why it matters for the current task, and how they should respond. \textbf{We therefore argue for a shift from text-centered uncertainty cues to user uncertainty management}: supporting how users incorporate uncertainty into situated task practices. U-Lens instantiates this shift by organizing uncertainty support around interpretation, evaluation, and decision stages, rather than treating uncertainty support as the presentation of isolated text-level cues. Our user study further demonstrates the value of this shift: compared with a confidence-cue baseline, U-Lens reduced false-positive checks, improved handoff quality, lowered perceived workload, and provided stronger perceived support across all three stages of uncertainty management.

Below, we discuss (1) how this process view extends TMIM for human--LLM interaction and its design implications, (2) how it motivates a broader shift from supporting the answer itself to supporting users' sensemaking work with AI-generated answers, (3) what new agency and coverage tensions arise when systems support users' uncertainty-management workflows, and (4) the limitations and future directions of this work.

\subsection{From Text-Centered Uncertainty Cues to User-Centered Uncertainty Management in Human-LLM Interaction using TMIM}
Prior TMIM research has often examined technology-mediated uncertainty management around human-produced digital information, such as profiles, posts, health resources, and community discussions. In human-LLM interaction, however, the object of uncertainty shifts from human-produced information to AI-generated artifacts. Unlike human-produced digital information, an LLM response is not an aggregation of human-authored content, but a probabilistic, black-box synthesis generated through next-token prediction \cite{brown2020language, bommasani2021opportunities}. This changes the reliability problem: users are not only judging who produced the information or whether a source is credible, but also evaluating a generation process whose intermediate rationale is often not explicitly inspectable. As a result, uncertainty may be distributed across details, claims, and relations throughout the response. \textbf{We therefore extend TMIM from managing uncertainty about human-produced digital information to managing uncertainty embedded in AI-generated artifacts themselves.}

\textbf{For the interpretation stage}, our findings extend TMIM from recognizing uncertainty in identifiable traces of human-produced digital information to constructing uncertainty targets in AI-generated artifacts. In settings centered on human-produced digital information, uncertainty is often anchored in an identifiable post, profile, message, or community discussion \cite{tokunaga2014seeking,kuang2022offline,bai2024research,wang2026testing}. In long-form LLM responses, however, the questionable object and its boundaries are not pre-given; potential uncertainty can emerge across varied textual structures and scales. Interpretation is therefore not only about recognizing bounded uncertainty, but also about turning distributed uncertainty at different granularities into concrete, inspectable targets.

This suggests that future uncertainty-support systems should be designed around target construction rather than only target marking. They should represent uncertainty targets as contextual, multi-granular interaction objects that users can inspect, compare, prioritize, and act on. The appropriate target granularity should depend on the task structure and verification practice, such as an entity in a biography, a causal relation in an explanation, or a parameter or function block in code generation.

\textbf{For the evaluation stage}, our findings extend TMIM by showing that uncertainty evaluation around AI-generated artifacts requires translating model-side generation signals into user-actionable factual and semantic hypotheses. In settings centered on human-produced digital information, evaluation often focuses on whether an author or source is knowledgeable, credible, or honest \cite{metzger2013credibility}. For AI-generated artifacts, reliability may be partially reflected in model-side signals, such as token probabilities or consistency \cite{lin2022teaching,jiang2021can,kuhn2023semantic}. However, these signals do not directly tell users whether a specific claim is factually grounded or needs verification \cite{min2023factscore}. Uncertainty support must therefore bridge low-level generation signals and high-level semantic validity by explaining why a target may be unreliable and what evidence would be needed to assess it. U-Lens instantiated this bridge by using generation-signal calibration and relation-extractor confidence to prioritize targets, then presenting each target as an evaluative hypothesis about what may be wrong and what evidence is needed. This made uncertainty more interpretable than confidence labels alone, which participants reported in the study as difficult to act on. Future systems should therefore treat model-side signals not as end-user explanations in themselves, but as inputs for higher-level semantic translation that better aligns with users’ cognitive practices.

Our findings further show that this support should be adaptive rather than uniform. For example, simple factual details may only require a concise concern, whereas unfamiliar topics or complex relational claims may require background context, terminology support, evidence boundaries, and possible error mechanisms. Future systems should therefore adapt uncertainty explanations to users' domain knowledge, task goals, and target complexity, helping users evaluate specific uncertainty targets rather than merely exposing generic model-side signals.

\textbf{For the decision stage}, our findings extend TMIM by showing that decisions around AI-generated artifacts are not only individual, directional choices about whether to seek or avoid information, but human-AI collaborative, operational decisions about how uncertainty should be handled. In prior TMIM settings, decision processes often concern a person's directional strategy, such as whether to seek, avoid, monitor, disclose, or share information \cite{brashers2001communication}. In long-form LLM interaction, however, users and systems must jointly shape more operational decisions, such as which part of an external source to inspect, what question to ask another model, or what evidence would be sufficient to resolve the uncertainty. This requires a clearer division of decision labor between system support and user judgment. In our study, U-Lens recommended verification actions and explained how each action could be carried out, while users retained responsibility for deciding whether to adopt the recommendation, conducting the check, and making the final verification judgment. In this way, U-Lens shortened participants' action-to-save time and improved handoff quality by narrowing and structuring the fact-checking work users needed to perform, without replacing their agency.

This suggests that future uncertainty-support systems should treat response guidance as collaborative planning rather than automated decision output.  Interfaces should make the decision labor visible by showing the recommended action, its rationale, and what remains for the user to judge, while allowing users to follow, adapt, reject, or defer the recommendation. In this way, AI can narrow open-ended uncertainty management work without taking over users' judgment and responsibility.

\subsection{Beyond Supporting the Answer: Designing Generative AI Outputs for User Sensemaking}
Generative AI interfaces have increasingly introduced support for users who face uncertainty about model-generated answers. Existing designs communicate uncertainty through confidence cues and verbal expressions \cite{kim2024m, xu2025confronting}, expose factuality scores and source attributions \cite{do2024facilitating,do2025highlight}, or use consistency across generated responses to surface reliability concerns \cite{cheng2024relic}. LLM interfaces also increasingly ground answers with supporting evidence, such as external sources and citations \cite{gao2023rarr,gao2023enabling,li2024citation}. These approaches enrich generated content with reliability signals and evidence, but much of this support still centers on the answer itself. Even with these signals and evidence, users still have substantial sensemaking work to do: they must determine how to interpret and work through the information, what deserves attention, and what to do next.

\textbf{Our findings suggest a shift in generative AI interface design from supporting the answer itself to supporting the user's work with the answer.} U-Lens illustrates this shift by starting from uncertainty signals in the generated content itself and presenting them as user-centered interaction objects. Rather than merely adding external evidence after generation, U-Lens focuses on how users progressively make sense of generative AI outputs in practice and provides support throughout this sensemaking and decision process. More broadly, we argue that \textbf{supporting user sensemaking should become an important design goal for future generative AI outputs, complementing the prevailing content-plus-evidence paradigm}.

This shift can also strengthen existing evidence-based LLM interfaces. First, when an interface presents many sources or citations, user-centered sensemaking support can reduce the substantial judgment work otherwise placed on users, especially those with limited domain expertise. Second, evidence-based interfaces have practical limits: external information may be incomplete, contradictory, unavailable for emerging or local contexts, or too costly to retrieve. In such cases, model-internal generation signals can complement external evidence by helping surface, organize, and prioritize potential concerns without requiring external retrieval.

This shift can apply across broader LLM use contexts beyond the biography, event explanation, and scientific explanation tasks examined in our study. In learning and sensemaking, citations may support individual concepts while leaving generated relationships unclear \cite{gao2023enabling,suh2023sensecape}. In technical and knowledge work, sources may still leave version-specific assumptions, parameter constraints, or contextual dependencies unresolved \cite{wang2025llms,yun2025generative}. In these settings, user-centered sensemaking support can complement content and evidence by helping users identify what remains uncertain, understand why it matters, and determine appropriate follow-up actions.

Beyond LLM interfaces, this perspective may also inform HCI systems for fact-checking, misinformation review, AI auditing, and content moderation. Existing systems provide valuable support through credibility indicators and source labels \cite{yaqub2020effects,heuer2022comparative}, checklists \cite{heuer2022comparative}, personalized misinformation judgments \cite{jahanbakhsh2023exploring}, in-place assessments \cite{jahanbakhsh2024browser}, and explainable automated fact-checking outputs \cite{warren2025show}. Rather than replacing these approaches, our framing suggests organizing such signals and assessments around the user's actual review workflow, reducing judgment burden while preserving users' control over final decisions.

\subsection{The Double Edge of User-Centered Uncertainty Management}
Prior HCI work on AI-assisted writing, sensemaking, and knowledge work shows a recurring tradeoff: systems can reduce users' cognitive work while introducing new coordination, control, and oversight demands \cite{kaur2020interpreting, buccinca2021trust, vasconcelos2023explanations}. U-Lens also reveals this tradeoff when shifting from text-centered cues to user-centered uncertainty management. Here, we elaborate on two main types of concerns.

\textbf{Information and oversight burden.} Prior work in human-centered AI has shown that adding explanations and supporting evidence can improve transparency but also require additional work from users \cite{abdul2020cogam}. Our findings extend this concern to uncertainty-management workflows. Some participants reported spending extra effort reading U-Lens explanations, ignoring unnecessary fields, and filtering information they did not need. Others described an oversight burden, where they actively monitored whether the system had omitted potentially important targets. Future systems should therefore design uncertainty support around this tradeoff: adapting explanation depth to users' task needs and effort constraints, preserving access to unmarked content, and clearly communicating what the system has and has not checked.

\textbf{Autonomy and coverage boundaries.} Prior work has examined when AI systems should support or replace user decision-making \cite{green2019principles, mei2026adapting}. U-Lens extends this concern to user uncertainty management. Even when users retain final responsibility for correctness judgments, the system can shape their judgment paths by selecting and prioritizing uncertainty targets. Focused checking may improve efficiency, but it can also lead users to over-attend to marked targets, under-read unmarked content, or treat the system's target set as the boundary of what matters. Over time, this could reinforce narrow attention patterns or confirmation bias \cite{sharma2024generative, liao2013beyond}. Future systems should preserve user autonomy across stages of uncertainty management: allowing users to adjust target selection, choose their own verification actions, and make final judgments based on the evidence. They should also allow users to switch between focused checking and full-response review.

These double-edged effects are also likely to be context-dependent. High-stakes or accuracy-critical tasks, such as healthcare or legal tasks, may require deeper explanations and stronger oversight, whereas low-stakes exploration may benefit from lighter, optional support. Future systems should calibrate uncertainty-management support to the task goal and downstream consequences, rather than assuming that more uncertainty support is always better.

\subsection{Limitations and Future Work}
This work has several limitations. First, our studies focused on long-form LLM responses in biography, event explanation, and scientific explanation tasks. While these materials enabled controlled comparison, they do not cover the full range of uncertainty in everyday LLM use. Also, U-Lens primarily addresses factual uncertainty and does not yet support more subjective uncertainty, such as moral standards, value judgments, or normative trade-offs. Future work should examine broader domains and task contexts.
Second, U-Lens should not be understood as a complete factual verification system. It provides users with entry points for checking, but its suggested targets and actions may still miss important errors, ambiguous assumptions, or underspecified claims. Future systems should combine uncertainty-management interfaces with evidence-oriented support, such as retrieval, source comparison, and post-verification correction.
Third, our evaluation used a controlled handoff-verification scenario in a lab study rather than sustained real-world use. This design helped isolate U-Lens's effects, but it simplified how people use LLM outputs through iterative practices over time. Future studies should examine longitudinal and open-ended settings that include users' own sources, verification habits, and LLM-use routines.
Finally, U-Lens implements a limited set of design elements, including target types, explanation fields, and response actions. This scope was sufficient to evaluate our core claim of user-centered uncertainty management, but future work could support more adaptive functionality, such as personalized verification preferences, adjustable ranking criteria, and movement between focused checking and full-response review. 

%% file: data/chap8.tex
\section{Conclusion}
This paper investigated how interactive systems can support user uncertainty management in long-form LLM responses, extending and operationalizing TMIM for AI-generated information. Through U-Lens and its evaluation, we show that uncertainty support can move beyond isolated cues to scaffold how users make sense of and work with uncertain AI-generated content, improving verification efficiency and handoff quality while reducing workload. More broadly, we argue that future generative AI interfaces should shift from supporting the answer itself to supporting the user's work with the answer, making user sensemaking a critical design goal alongside content generation and evidence provision.

%% file: data/appendix.tex
\newpage

\section{Formative Study Details}

\subsection{Participant Demographics}
\label{section: Participant Demographics}
\autoref{tab:formative_study_demographic} provides detailed demographic information for each participant in the formative study.

\begin{table}[htbp]
\caption{Demographic details of participants in the formative study.}
\label{tab:formative_study_demographic}
\small
\begin{tabular}{ccccc}
\toprule[1pt]
\textbf{ID} & \textbf{Gender} & \textbf{Age} & \textbf{Education} & \textbf{LLM Experience} \\ \hline
P1          & Female          & 21           & Bachelor's         & \textgreater{}2 years   \\ \hline
P2          & Female          & 23           & Master's           & 1-2 years               \\ \hline
P3          & Female          & 25           & PhD                & \textgreater{}2 years   \\ \hline
P4          & Female          & 21           & Bachelor's         & 1-2 years               \\ \hline
P5          & Male            & 21           & Bachelor's         & 1-2 years               \\ \hline
P6          & Female          & 23           & Master's           & \textgreater{}2 years   \\ \hline
P7          & Male            & 27           & Master's           & \textgreater{}2 years   \\ \hline
P8          & Male            & 25           & PhD                & 1-2 years               \\ \hline
P9          & Male            & 22           & Master's           & \textgreater{}2 years   \\ \hline
P10         & Female          & 21           & Bachelor's         & \textgreater{}2 years   \\ \hline
P11         & Female          & 25           & PhD                & 1-2 years               \\ \hline
P12         & Female          & 22           & Bachelor's         & \textgreater{}2 years   \\ \bottomrule[1pt]
\end{tabular}
\end{table}

\subsection{Study Materials}
\label{section: Study Materials}
We provide representative examples of the study materials used in our formative study. These materials cover three task types: biography generation, event explanation, and scientific explanation.

\noindent\rule{\linewidth}{0.2pt}

\textbf{Task type:} biography generation

\textbf{Prompt:} \textit{Tell me a bio of Zhang Yiming.}

\textbf{LLM output (with factual errors):}
\begin{tcolorbox}[
    breakable,
    colback=gray!8,
    colframe=gray!25,
    boxrule=0.1pt,
    arc=3mm,
    left=4mm,
    right=4mm,
    top=2mm,
    bottom=2mm,
    before skip=8pt,
    after skip=12pt,
]
\small
Zhang Yiming, born in April 1983, is a prominent Chinese entrepreneur and investor, and the founder and Chief Executive Officer (CEO) of ByteDance.

Zhang was born in Fujian, China. He graduated from the School of Software at Nankai University with a bachelor’s degree in computer science. During his university years, he demonstrated outstanding programming skills and innovative thinking, developing several software projects.

In 2006, after graduating from Nankai University, Zhang joined Innovation Works in Shenzhen, where he worked in software development and product management. In 2009, he founded his first company, ``99fang,'' a real estate information website that later became a leading platform in China for property information services.

In 2012, Zhang founded ByteDance. The company initially focused on developing a news recommendation app called Toutiao. Leveraging its distinctive algorithmic technology and personalized recommendation system, Toutiao quickly achieved significant success in the Chinese market and later expanded globally.

ByteDance subsequently launched several globally recognized mobile applications, including TikTok, Xigua Video, Neihan Duanzi, and Xiaohongshu. These products have gained massive user bases and substantial market influence across multiple countries and regions. ByteDance has since grown into one of the most influential technology companies in the world.

Zhang Yiming is known for his low-profile, pragmatic style and strong emphasis on technological innovation and user experience. He has consistently upheld the mission of “making information more accessible,” promoting the deep integration of artificial intelligence, big data, and content distribution technologies. He is also an investor in multiple technology companies, supporting the incubation and development of various innovative projects.

As of today, Zhang remains a core leader of ByteDance. His entrepreneurial journey and leadership style continue to provide valuable insights for the technology industry in China and around the world.

\end{tcolorbox}

\noindent\rule{\linewidth}{0.2pt}

\textbf{Task type:} event explanation

\textbf{Prompt:} \textit{Tell me a history of the Apollo 13 mission.}

\textbf{LLM output (with factual errors):}
\begin{tcolorbox}[
    breakable,
    colback=gray!8,
    colframe=gray!25,
    boxrule=0.1pt,
    arc=3mm,
    left=4mm,
    right=4mm,
    top=2mm,
    bottom=2mm,
    before skip=8pt,
    after skip=12pt,
]
\small
Apollo 13 was the third crewed lunar landing mission launched by the NASA in 1970. Originally intended to land astronauts on the Moon, the mission was aborted after a critical malfunction. It ultimately became known as a “successful failure,” demonstrating human ingenuity and cooperation in the face of crisis.

Mission Background

Apollo 13 was the third lunar mission of the Apollo program, codenamed “Salvage.” The crew consisted of Commander Jim Lovell, Lunar Module Pilot Fred Haise, and Command Module Pilot Jack Swigert. The mission’s primary objective was to land astronauts on the Moon and return them safely to Earth.

The Accident

On April 13, 1970, during the flight, Apollo 13 suffered a catastrophic explosion in an oxygen tank located in the service module. The blast caused a loss of oxygen supply, damage to electrical systems, and failure of life-support functions. The situation nearly resulted in the loss of the crew, but through the combined efforts of the astronauts and mission control, they managed to survive and return safely.

Astronaut Response

Following the explosion, the astronauts immediately took emergency measures, shutting down damaged systems and activating backup procedures. They adopted an extreme power-saving mode, using the Lunar Module as a lifeboat (“Lunar Module as lifeboat”) to sustain life-support functions while maintaining communication with mission control.

Support from Mission Control

NASA’s ground control team in Houston worked closely with the crew to devise detailed emergency procedures, including conserving power, managing water and waste, and rationing food. Their coordinated efforts enabled the astronauts to endure several days of hardship and safely return to Earth on April 17, 1970, landing in the South Pacific Ocean.

Significance and Impact

Although Apollo 13 did not achieve a lunar landing, it became one of the most significant missions in spaceflight history. It highlighted human resilience, calm decision-making, and teamwork under extreme conditions, and contributed to improvements in space technology and safety. The mission later inspired the film Apollo 13, which vividly portrays the tension and struggle of the event.

Conclusion

Apollo 13 was not merely a failed lunar mission, but a triumph of teamwork, innovation, and the human spirit. It serves as a reminder that even in the most challenging circumstances, science, technology, and human ingenuity can overcome adversity.

\end{tcolorbox}

\noindent\rule{\linewidth}{0.2pt}

\textbf{Task type:} scientific explanation

\textbf{Prompt:} \textit{Tell me an explanation of the Kepler Space Telescope.}

\textbf{LLM output (with factual errors):}
\begin{tcolorbox}[
    breakable,
    colback=gray!8,
    colframe=gray!25,
    boxrule=0.1pt,
    arc=3mm,
    left=4mm,
    right=4mm,
    top=2mm,
    bottom=2mm,
    before skip=8pt,
    after skip=12pt,
]
\small
The Kepler Space Telescope was a space telescope launched by the NASA specifically to detect exoplanets. It was launched on March 7, 2009.

Primary Function

The main mission of the Kepler telescope was to detect planets outside the Solar System using the “Transit Method.” When a planet passes between its host star and the Earth, it briefly blocks a small portion of the star’s light, causing a slight dip in brightness. Kepler could detect these changes through high-precision photometric measurements, thereby determining the presence of planets.

Technical Features

High Sensitivity: Kepler was equipped with a highly precise photometer capable of detecting extremely small variations in brightness, even identifying Earth-sized planets.

Wide Field of View: Its observational field covered about 100 square degrees of the sky, roughly equivalent to about one-quarter of the sky visible from Earth.

Long-Term Observation: It operated in space for over nine years and discovered thousands of exoplanets.

Major Discoveries

Kepler discovered many Earth-sized planets, including some located within the “habitable zone” of their stars (regions where liquid water could potentially exist).
The vast number of exoplanets it identified greatly advanced humanity’s understanding of whether other forms of life might exist in the universe.

End of Mission

Due to failures in its reaction wheels (critical components used to maintain orientation), Kepler was officially declared retired in 2018. However, NASA later repurposed its remaining capabilities into a follow-up mission known as K2, allowing it to continue astronomical observations.

Summary

The Kepler Space Telescope has been a crucial tool in humanity’s exploration of exoplanets and the search for extraterrestrial life. Its discoveries have significantly expanded our understanding of planetary systems in the universe.
\end{tcolorbox}

\section{U-Lens Target Calibrator Details}
\label{appendix:calibrator-details}

We trained the U-Lens target calibrator as an offline target-level model. The calibration data were constructed from 150 long-form Qwen3-8B responses across biography generation, event explanation, and scientific explanation tasks, resulting in 7059 target-level records for model fitting and validation: 4698 token-level targets, 1975 sentence-level targets, and 386 segment-level targets. We split the curated data by generated response, holding out 30 responses for model testing and using the remaining 120 responses for model training, comparison and feature selection.

Each target was encoded with 19 generation-time signals and \texttt{target\_type}. Within the training responses, we compared model families using the full signal set under 5-fold cross-validation, using AUROC to assess whether higher-scored targets tended to be riskier. \autoref{tab:target-calibrator-model-results} summarizes the comparison. The balanced L1 and L2 logistic models and the small MLP formed the strongest group; we retained the class-balanced L2 logistic model because it provided comparable AUROC and a simple probabilistic form for deployment.

Within the L2 logistic family, we performed forward wrapper selection on the training responses, keeping \texttt{target\_type} in every candidate model. At each step, we added the generation-signal feature with the highest training-set 5-fold cross-validated AUROC. We then applied the one-standard-error rule and retained the smallest prefix whose mean AUROC was within one standard error of the best prefix. The final six-signal set contains:
\begin{itemize}[leftmargin=*,nosep]
\item \texttt{max\_token\_uncertainty}, \texttt{max\_uncertainty\_rank\_percentile}
\item \texttt{token\_count\_log}, \texttt{span\_char\_length\_log},
\item \texttt{mean\_margin}, \texttt{mean\_uncertainty\_rank\_percentile}
\end{itemize}

\autoref{tab:selected-calibrator-target-type-results} reports the final six-signal calibrator on the test data. U-Lens uses these scores as prioritization cues rather than factuality judgments.

For target $i$, let $x_i$ denote the selected signal vector and $g_i$ denote the target type. The calibrator estimates the probability that the target is supported:
\[
P_\theta(y_i=1 \mid x_i,g_i)=\sigma(\beta_0+\beta^\top \tilde{x}_i+\gamma_{g_i}),
\]
where $y_i=1$ indicates a supported target, $\tilde{x}_i$ is the standardized feature vector, $\gamma_{g_i}$ is the target-type term, and $\sigma(\cdot)$ is the logistic function. Parameters were fit by minimizing class-balanced binary cross-entropy with L2 regularization on all non-intercept coefficients:
\[
\min_{\beta_0,\beta,\gamma}
-\sum_{i=1}^{N} w_{y_i}
\left[
y_i \log p_i + (1-y_i)\log(1-p_i)
\right]
+
\lambda \left(\lVert \beta \rVert_2^2+\lVert \gamma \rVert_2^2\right),
\]
where $p_i=P_\theta(y_i=1 \mid x_i,g_i)$ and $w_{y_i}$ is the class-balancing weight.

U-Lens uses the complement as the target priority score:
\[
S_i=1-P_\theta(y_i=1 \mid x_i,g_i).
\]

\begin{table*}[htbp]
\centering
\small
\caption{Target-calibrator model comparison on training responses.}
\label{tab:target-calibrator-model-results}
\begin{tabular}{p{0.3\linewidth}r}
\toprule
Model & AUROC \\
\midrule
Heuristic signal score & .677 $\pm$ .017 \\
Logistic regression, L1 balanced & .791 $\pm$ .016 \\
Logistic regression, L2 balanced & .790 $\pm$ .015 \\
Type-specific logistic, L2 balanced & .759 $\pm$ .013 \\
Gradient boosting & .775 $\pm$ .014 \\
Small MLP, balanced & .789 $\pm$ .012 \\
Histogram gradient boosting & .768 $\pm$ .011 \\
Extra trees, balanced & .749 $\pm$ .015 \\
Random forest, balanced & .759 $\pm$ .013 \\
\bottomrule
\end{tabular}
\end{table*}

\begin{table*}[htbp]
\centering
\small
\caption{Final calibrator performance on the test data by target type. The all-unit row is computed by pooling all unit-based targets.}
\label{tab:selected-calibrator-target-type-results}
\begin{tabular}{p{0.2\linewidth}rr}
\toprule
Target type & Risk prev. & AUROC \\
\midrule
All unit targets  & .192 & .726 \\
Token-level  & .144 & .717 \\
Sentence-level  & .248 & .650 \\
Segment-level  & .467 & .754 \\
\bottomrule
\end{tabular}
\end{table*}
